\documentclass[12pt]{article}
\makeatletter
\catcode`\@=11
\@addtoreset{equation}{section}

\makeatother

\baselineskip=\normalbaselineskip

\usepackage{mathrsfs,amsbsy,amssymb,latexsym,amsfonts,amsmath}
\usepackage{graphicx,color}

\allowdisplaybreaks[4]
\renewcommand{\thefootnote}{\arabic{footnote}}
\def\rmd{{\mathrm{d}}}
\def\ii{\operatorname{i}}
\newcommand{\Exp}[1]{\, \operatorname{e}^{\, #1}}

\begin{document}

\begin{titlepage}
\renewcommand{\thefootnote}{\fnsymbol{footnote}}

\begin{flushright}
\parbox{3.5cm}
{KEK-TH-1547}

\parbox{3.5cm}
{KUNS-2401}

\parbox{3.5cm}
{MISC-2012-09}
\end{flushright}

\vspace*{1.0cm}

\begin{center}
{\Large \bf Rotating string in doubled geometry \\ with generalized isometries}%
\end{center}
\vspace{1.0cm}

\centerline{
{Toru Kikuchi}%
\footnote{E-mail address: 
kikut@post.kek.jp}, 
{Takashi Okada}%
\footnote{E-mail address: 
okada@gauge.scphys.kyoto-u.ac.jp} 
and
{Yuho Sakatani}%
\footnote{E-mail address: 
yuho@cc.kyoto-su.ac.jp}%
}

\vspace{0.2cm}

\begin{center}

${}^\ast${\it Theory Center, High Energy Accelerator Research Organization (KEK)\\
Tsukuba, Ibaraki 305-0801, Japan}

${}^\dagger${\it Department of Physics, Kyoto University \\ 
Kyoto 606-8502, Japan\\}

${}^\ddagger${\it Maskawa Institute for Science and Culture, Kyoto Sangyo University,\\
Kyoto 603-8555, Japan\\}

\end{center}
\vspace*{1cm}
\begin{abstract}
In this paper, we first construct a globally well-defined non-geometric background which contains several branes in type II string theory compactified on a 7-torus. One of these branes is called $5^2_2$, which is a codimension-2 object and has a non-trivial monodromy given by a $T$-duality transformation. The geometry near the $5^2_2$-brane is shown to approach the non-geometric background constructed in arXiv:1004.2521. We then construct the solution of a fundamental string rotating along a non-trivial cycle in the $5^2_2$ background. Although the background is not axisymmetric in the usual sense, we show that it is actually axisymmetric as a doubled geometry by explicitly finding a generalized Killing vector. We perform a generalized coordinate transformation into a system where the generalized isometry is manifest, and show that the winding and momentum charges of the string solution is explicitly conserved in that system.
\end{abstract}
\thispagestyle{empty}
\end{titlepage}

\tableofcontents

\setlength{\parskip}{0.3\baselineskip}

\setcounter{footnote}{0}

\section{Introduction}

In string theory, there exist non-geometric backgrounds, 
called \emph{$U$-folds} \cite{Hull:2004in}. 
The $U$-folds are backgrounds in which 
the background fields on any two coordinate patches are related 
by transition functions which generally belong to the $U$-duality group. 
In other words, they are characterized by the property that 
as we go around a non-trivial cycle in the backgrounds, 
the background fields come back to themselves
only up to the $U$-duality transformations. 
Since the $U$-duality group is the symmetry of string theory, 
the $U$-folds  naturally appear as the backgrounds of string theory. 
The $U$-duality group contains the $T$-duality group and the $S$-duality group as its subgroups, 
and $T$-folds and $S$-folds are defined in a similar manner. 

In spite of the unusual properties, there are many $U$-fold backgrounds in type II string theory. 
The simplest example is the ${\rm D7}$-brane background. 
The circular integral of the Ramond-Ramond (RR) flux $\rmd C_0$ over a circle $S$ 
enclosing the ${\rm D7}$-brane is given by 
$\int_S \rmd C_0=C_0\bigr\rvert_{\theta=2\pi}-C_0\bigr\rvert_{\theta=0}=1$, 
so the value of $C_0$ is shifted by $1$ as we go around the ${\rm D7}$-brane. 
In other words, the ${\rm D7}$-brane background has the non-trivial monodromy 
$C_0\rightarrow C_0+1$, which is nothing but an element 
of $S$-duality group ${\rm SL}(2,\mathbb{Z})$. 
In this sense, the ${\rm D7}$-brane background is an $S$-fold. 

\begin{figure}[t]
\begin{center}
\includegraphics[width=12cm]{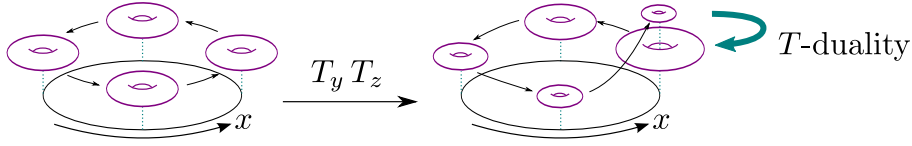}
\caption{A torus parameterized by $y$ and $z$ is fibered on the base $S^1$ parameterized by $x$. After taking $T$-dualities along $y$- and $z$-directions ($T_y$ and $T_z$), the torus becomes non-trivially fibered on the $S^1$, whose transition function is given by $T$-duality.}
\label{fig:t3}
\end{center}
\end{figure}

One might not be surprised at the ${\rm D7}$-brane example 
since the transformation connecting $C_0$ and $C_0+1$ 
is merely a gauge transformation. 
Things get more interesting when we consider the case of $T$-folds. 
A simple example of $T$-folds \cite{Hull:2004in} can be constructed as follows 
(although it is not a solution of supergravity). 
Consider a flat 3-torus $T^3$ with constant $3$-form $H$-flux on it. 
We regard the 3-torus as a $T^2$-fibration (spanned by coordinates $y$ and $z$) 
over a base $S^1$ with coordinate $x$ and periodicity $x\sim x+1$ (see Figure \ref{fig:t3}).
In this background, if we parameterize the flux as $H=\rmd B^{(2)}=N \, \rmd x \wedge \rmd y \wedge \rmd z$ 
with $N$ an arbitrary constant, 
the $2$-form $B^{(2)}$ is given by $B^{(2)}= N\, x \, \rmd y \wedge \rmd z$ in a specific gauge.
Then, as we go along the base circle, $x\rightarrow x+1$, 
the $B^{(2)}$ field shifts by $B^{(2)}\rightarrow B^{(2)}+N\, \rmd y \wedge \rmd z$.
This shift is again purely a gauge transformation. 
The shift becomes non-trivial if we perform the $T$-duality transformations along $y$- and $z$-directions.
After the $T$-duality transformations the resulting geometry takes the following form:
\begin{align}
 \rmd \ell^2 = \rmd x^2 +\frac{\rmd y^2+\rmd z^2}{1+N^2 x^2}\,,\qquad
 B^{(2)} = \frac{Nx}{1+N^2 x^2}\, \rmd y\wedge \rmd z\,.
\end{align}
We can easily check that some components of the Riemann curvature tensor are aperiodic 
and do not come back to their original values as we go along the base circle $x\rightarrow x+1$ 
(e.g., the Ricci scalar $R=-2N^2\,(5 N^2 x^2-2)/(1+N^2 x^2)^2$ is aperiodic). 
This aperiodicity arises from the fact that the $T$-duality transformations mix the 
$B$-field (which is aperiodic in the original frame) and the metric together. 

\begin{figure}[t]
\begin{center}
\includegraphics[width=16cm]{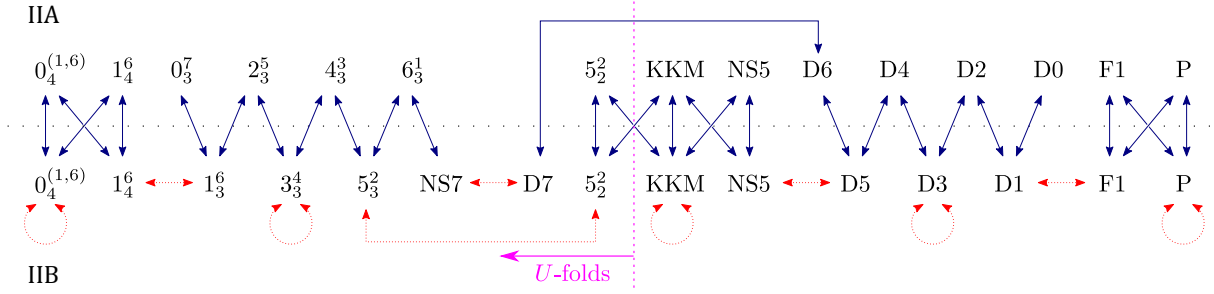}
\caption{The dotted horizontal line separates Type IIA theory (above) and Type IIB theory (below). Branes connected with a blue line are related by a $T$-duality to each other, and those connected with a red line are related by an $S$-duality to each other. Backgrounds of the branes in the left of the dashed vertical line are $U$-folds.}
\label{fig:web}
\end{center}
\end{figure}

In type II string theory compactified on a 7-torus, we can obtain many kinds of $U$-folds 
by taking dualities as shown in the left side of Figure \ref{fig:web}. 
Note that all of these $U$-folds correspond to codimension-2 branes.
Here, the notation such as $b_n^c$ or $b_n^{(d,c)}$ indicates that the mass 
of the brane depends linearly, quadratically and cubically on the lengths 
of $b$, $c$ and $d$ of the compactified longitudinal directions, respectively, 
and that the mass is proportional to $g_s^{-n}$ where $g_s$ is the string coupling constant. 
We note that there are, in general, several directions 
along which we can perform a $T$-duality transformation. 
For example, for a ${\rm KKM}(56789,4)$, taking $T_3$ leads to a $5_2^2 (56789,34)$, 
taking $T_4$ to an ${\rm NS}5(56789)$, and taking $T_5$ to a ${\rm KKM}(56789,4)$. 
Here, the notation of $5_2^2 (56789,34)$ indicates that 
the mass dependence of the $5_2^2$ is given by $M=R_5\, R_6\, R_7\, R_8\, R_9\, (R_3\, R_4)^2/g_s^2 l_s^9$, 
and that of ${\rm KKM}(56789,4)$ (${\rm KKM}$ can also be denoted by $5_2^1$) indicates that 
the mass is given by $M= R_5\, R_6\, R_7\, R_8\, R_9\, R_4^2/g_s^2 l_s^8$, 
where $R_n$ ($n=3,\dotsc 9$) are radii of the 7-torus. 
Branes whose masses have such unusual dependence on the compactification 
radii and the string coupling constant are sometimes called \emph{exotic branes} 
\cite{Elitzur:1997zn,Blau:1997du,Hull:1997kb,Obers:1997kk,Obers:1998fb}. 
In particular, the $5_2^2$-background is known to be a $T$-fold \cite{deBoer:2010ud}, 
whose explicit form will be displayed later in Eq.~\eqref{522sol}.

Moreover, at least in supergravity theory, 
we can construct additional $U$-folds, 
which would correspond to the bound states of several 7-branes. 
The reason for considering these bound states is that 
we generally cannot put only a single kind of 7-branes 
in order for the resulting background to be well-defined 
\cite{Bergshoeff:2006jj,Bergshoeff:2008zz}. 
We must put branes with other charges (monodromies) 
which are not related  each other via $U$-duality.  
For example, a ${\rm D7}$-brane generally must accompany branes 
such as an \emph{$S$-brane} \cite{Bergshoeff:2006jj}, 
whose monodromy is the $S$-duality transformation, $\lambda\rightarrow -1/\lambda$, 
where $\lambda$ is axion-dilaton field. 
Likewise, the $5^2_2$-brane must accompany other kinds of branes, 
one of which we here call a \emph{$T_{34}$-brane}, whose monodromy is given by $T_3\, T_4$. 
We will see the detail in section \ref{sec:522}.

Given the above non-geometric backgrounds, it is then a natural step 
to study what happens when we put a probe on such backgrounds.  
When we drag the probe with charge ${\bf Z}$ around a brane with monodromy $\Omega$, 
the probe charge is affected by the monodromy of the brane on the way around it. 
After coming full circle around the brane, the probe charge eventually becomes
\begin{align}
 {\bf  Z} \rightarrow \Omega^{-1} \cdot {\bf Z}\,. 
\end{align}
One of our motivations in this paper is to investigate how the probe charge, ${\bf  Z}$, 
would vary in the course of traveling along non-trivial cycles. 

The codimension-2 objects with non-trivial monodromies  appear not only in string theory 
but also in ordinary field theories as vortices. 
It was argued long time ago that certain gauge theories admit the existence of a vortex solution, 
called an \emph{Alice string} \cite{Schwarz:1982ec}. 
It has the property that a particle with charge of unbroken gauge symmetry 
changes the sign of its charge  when it goes around the Alice string;
$e \rightarrow -e$.
Such an exotic property of the Alice string has been applied into various fields 
in physics such as the cosmology based on Grand Unified Theories 
\cite{Brekke:1991ap,BenMenahem:1992fe,Kobayashi:2010na}, 
nematic liquid crystals and superfluid 
in condensed matter physics \cite{Stephen:1974ur,Wright:1989zz} 
(see also \cite{Preskill,Vilenkin,Volovik} and references therein).

In this paper, we put a fundamental string as a probe in some $U$-folds such as $5_2^2$, 
and examine how the string behaves in such non-geometric backgrounds. 
In section \ref{sec:522}, we construct a $U$-fold background, which contains $5_2^2$,  
by $T$-dualizing the 7-brane solution constructed in \cite{Greene:1989ya}. 
We compare our construction of the $5_2^2$ background 
with that already known in \cite{deBoer:2010ud}. 
Our construction gives some physical understanding, 
which was not made clear in \cite{deBoer:2010ud}. 
We also study the monodromies of the background. 
Next, we construct fundamental string solutions rotating around $5_2^2$ 
in section \ref{string_solutions} 
(and that around $T_{34}$-brane in appendix \ref{sec:T34_string_solution}), 
and study how the momentum and winding charges of the string, ${\bf Z}^I$, 
transform under the action of the monodromy matrix of $5_2^2$. 
In section \ref{sec:DFT}, we review the double field theory 
\cite{Hull:2009mi,Hull:2009zb,Hohm:2010jy,Hohm:2010pp} 
and the definition of the \emph{generalized Lie derivative}. 
Although the $5_2^2$ is not axisymmetric due to the action of the monodromy matrix, 
it is shown to be axisymmetric if we regard the background as a \emph{doubled geometry}. 
Indeed, we find a \emph{generalized Killing field} generating the axial rotation, 
along which the generalized metric is invariant. 
For the string solution given in section \ref{string_solutions}, 
the charge  ${\bf Z}^I$ changes by 
${\bf Z}^I\rightarrow \bigl(\Omega_{\theta=2\pi}^{-1} \cdot {\bf Z} \bigr)^I$ 
with $\Omega_{\theta=2\pi}$ a constant matrix, after the string comes full circle. 
However, if we move to a coordinate system where the generalized isometry is manifest, then the charge density vector, which we denote by $Z^I$, is constant under the time evolution, that is, $\partial_\tau Z^I =0$. 
Conclusion and discussions are given in section \ref{sec:conclusion}. 

\section{$5^2_2$ solutions}\label{sec:522}

\subsection{$5_2^2$ from Kaluza-Klein monopole}\label{sec:fromKKM}

We first review the construction of the $5_2^2$ solution 
from the smeared Kaluza-Klein monopole (KKM) background 
by taking $T$-duality in a direction transverse to the (smeared) KKM \cite{deBoer:2010ud}. 
The solution of multiple ${\rm KKMs}(56789,4)$ is given by
\begin{align}
 &\rmd s^2 = \rmd x^2_{056789} + H\, \rmd x_{123}^2 + H^{-1}\, (\rmd x^4 + \omega)^2\,,\nonumber\\
 &H = 1+ \sum_n \frac{R_4}{2 \lvert\vec{x} - \vec{x}_n\rvert}\,,\quad 
 \Exp{2\phi} = 1\,, \quad \rmd \omega = \star_{3} \rmd H \,, \label{kkm}
\end{align}
where $\vec{x}_n \in {\mathbb{R}}^3_{123}$ in $H$ represent 
the centers of the KKMs in ${\mathbb{R}}^3_{123}$ 
and $\rmd x^2_{p\cdots q}$ represents the line element of the flat metric 
(e.g., $\rmd x^2_{056789}\equiv -(\rmd x^0)^2+(\rmd x^5)^2+\cdots +(\rmd x^9)^2$).
The periodicity of the 6-torus, over which the KKMs are wrapped, 
is given by $x^i \sim x^i+2\pi R_i$ ($i=4,\dotsc,9$). 
After smearing KKMs along $x^3$ with the interval $2\pi \tilde{R}_3$, 
the harmonic function $H$ is given by
\begin{align}
 H=& 1+\frac{1}{2\pi\tilde{R}_3}\int^L_{-L}\rmd \bar{x}^3\frac{R_4}{2\sqrt{r^2+(x^3-\bar{x}^3)^2}}
 \nonumber \\
  = & 1+\gamma\, \log \bigl[(L+ \sqrt{r^2+L^2})/r\bigr] \,,
\end{align}
where we have defined $r\equiv \sqrt{x_1^2 +x_2^2}$ and $\gamma \equiv R_4/(2\pi \tilde{R}_3)$. 
The integral is logarithmically divergent, 
and so we have introduced an IR cutoff $L$ for the integration. 
By introducing a ``renormalization'' scale $\mu$, 
we separate the divergence in $H$ as follows:\footnote{The separation between the divergent part and the finite part in $H$ is not unique, but this is not important for the discussion.}
\begin{align}
 H = h+\gamma\, \log (\mu/r)\,, \label{cutoff_mu}
\end{align}
where $h$ is the divergent term in the limit of $L \rightarrow \infty$. 
Then, $\omega$ in \eqref{kkm} is given by, for example, $\omega=-\gamma\, \theta\, \rmd x^3$, 
where $\theta$ is the polar angle in $\mathbb{R}^2_{12}$.
The aperiodicity about $\theta$ can be absorbed by a mere diffeomorphism 
$x^4 \rightarrow x^4-2\pi\gamma x^3$.  

We obtain the $5^2_2$ solution by taking $T$-duality of the smeared KKM along $x^3$.  
The result is given as follows \cite{deBoer:2010ud}:
\begin{align}
 \rmd s^2 &= H\, (\rmd r^2 + r^2 \rmd\theta^2) + H\, K^{-1}\,\rmd x_{34}^2 + \rmd x_{056789}^2 \,, 
 \nonumber\\
 B^{(2)} &= B_{34}\, \rmd x^3\wedge \rmd x^4 
          = - K^{-1}\,\gamma\,\theta \, \rmd x^3\wedge \rmd x^4 \,, 
\label{522sol}\\
 \Exp{2\phi} &= H\,K^{-1}\,,\quad K \equiv H^2 + \gamma^2\, \theta^2\,. \nonumber
\end{align}
Here, $\theta$-dependence appears in the metric on the internal 2-torus 
spanned by $x^3$ and $x^4$, 
and these components are aperiodic under $\theta\rightarrow \theta+2\pi$. 
It is then convenient to introduce the \emph{generalized metric} in the $x^3$-$x^4$ space 
by the following $4\times 4$ matrix:
\begin{align}
 (\mathcal{H}_{AB})\equiv \begin{pmatrix}
  G^{-1} & G^{-1}\,B \\
  -B\, G^{-1} & G-B\, G^{-1}\,B 
 \end{pmatrix}= \frac{1}{H} \begin{pmatrix}
 K\, {\bf 1}& -\gamma\,\theta \,\boldsymbol{\epsilon} \\
 \gamma\theta\,\boldsymbol{\epsilon} & {\bf 1}
\end{pmatrix}\,, \label{gen_metric}
\end{align}
where we defined $G\equiv \Bigl(\begin{smallmatrix}
  G_{33}& G_{34} \cr
  G_{43} & G_{44}
\end{smallmatrix}\Bigr)$, 
$B\equiv \Bigl(\begin{smallmatrix}
  B_{33}& B_{34} \cr
  B_{43} & B_{44}
\end{smallmatrix}\Bigr)$, 
${\bf 1}\equiv \Bigl(\begin{smallmatrix}
  1& 0 \cr
  0& 1
\end{smallmatrix}\Bigr)$ and 
$\boldsymbol{\epsilon}\equiv \Bigl(\begin{smallmatrix}
  0& 1 \cr
  -1& 0
\end{smallmatrix}\Bigr)$. 
We can easily check that the generalized metric of the $5^2_2$ at each $\theta$ 
is related to its value at $\theta=0$ by an ${\rm O}(2,2,\mathbb{R})$ transformation as follows:
\begin{align}
 \mathcal{H}(\theta)=\Omega_\theta^{\rm T}\,\mathcal{H}(\theta=0)\,\Omega_\theta \,, \qquad
 \Omega_\theta=
 \begin{pmatrix}
 {\bf 1} & 0 \\
 \gamma\theta\,\boldsymbol{\epsilon} & {\bf 1}
 \end{pmatrix}
 \in {\rm O}(2,2,\mathbb{R})\,.
\label{H_OHO}
\end{align}
We see that the monodromy matrix of $5^2_2$, $\Omega_{\theta=2\pi}$, 
takes the form neither of diffeomorphisms, 
$\Omega_{\rm diff}=\begin{scriptsize}\begin{pmatrix} X^{\rm T} & 0 \cr 0 & X^{-1}\end{pmatrix}\end{scriptsize}$ with $X\in GL(2,\mathbb{R})$, 
nor of the gauge transformations of $B^{(2)}$, 
$\Omega_{\rm gauge}=\begin{scriptsize}\begin{pmatrix} {\bf 1} & A \cr 0 & {\bf 1} \end{pmatrix}\end{scriptsize}$ with $A$ an antisymmetric matrix.  
It is a more general type of transformation 
which non-trivially mixes the metric tensor and the Kalb-Ramond field. 

An unsatisfactory point 
about the above $5^2_2$ solution obtained from the smeared KKM via $T$-duality is that
the solution is not globally defined 
and is valid only for $0<r<r_c$ with $r_c\equiv \mu \Exp{\frac{h}{\gamma}}$.
Outside the domain, $H$ given in Eq.~\eqref{cutoff_mu} becomes negative, 
which is obviously not allowed. 
In particular, the curvature scalar $R$ diverges at $r=0$ and $r=r_c$, 
though in the intermediate region 
it will be sufficiently small to neglect any higher curvature corrections. 
The appearance of the cutoff, $r=r_c$, 
originates from introducing the ``renormalization'' parameter $\mu$, 
whose physical meaning is not clear, in the smearing procedure.
We will clarify below that the cutoff parameter is to be interpreted 
as the distance of the $5^2_2$ from the other neighboring 7-branes. 

\subsection{$5^2_2$ from D7-brane}\label{sec:fromD7}

The $5_2^2$ solution given in Eq.~\eqref{522sol} is not globally 
well-defined owing to the cutoff at $r=r_c$, where the curvature diverges. 
In this subsection, we construct a globally well-defined $5_2^2$ solution 
by applying the following duality chain to the ${\rm D7}$-brane solution \cite{Greene:1989ya,Bergshoeff:2006jj}:
\begin{align}
 {\rm D7}(3456789) \overset{S}{\rightarrow}
 {\rm NS7}(3456789) \overset{T_3}{\rightarrow}
 6^1_3 (456789,3) \overset{T_4}{\rightarrow}
 5^2_3(56789,34) \overset{S}{\rightarrow}
 5^2_2(56789,34) \,.
\label{chain}
\end{align}
We note that in this construction, the $5_2^2$ background is constructed 
without any smearing procedure, 
and we will find that the physical meaning of the cutoff at $r=r_c$ is made clear. 

As is discussed in \cite{Greene:1989ya,Bergshoeff:2006jj}, 
we cannot consider the supergravity solution describing a single ${\rm D7}$-brane 
in order to make the energy density finite. 
The finite-energy ${\rm D7}$-brane solution must
contain several supersymmetric 7-branes (which have their own monodromies) 
other than a ${\rm D7}$-brane. 
In particular, in section \ref{sec:D7}, we construct the 7-brane solution 
which consists of a ${\rm D7}$-brane and the other two 7-branes. 
One of the others is an $S$-brane. 
By the duality transformations, $S\, T_3\, T_4\, S$, 
the ${\rm D7}$-brane is mapped to a $5_2^2$-brane 
while the $S$-brane is mapped to what we call a $T_{34}$-brane, 
whose monodromy is given by $T_3\,T_4\in {\rm O}(2,2;\mathbb{Z})$. 
We will check these monodromies in section \ref{522andI}. 

\subsubsection{D7-brane background}\label{sec:D7}

We first review the construction of the finite-energy 7-brane solution containing a ${\rm D7}$-brane 
presented in \cite{Greene:1989ya,Bergshoeff:2006jj}. 
A half-supersymmetric 7-branes background in the Einstein frame is generally written as follows;
\begin{align}
 &\rmd s^2 = \rmd x^2_{03456789}+ \lambda_2\, \lvert f\rvert^2\, \rmd z\, \rmd \bar{z} 
 \label{7_metric}\\
 &\lambda = \lambda(z)=\lambda_1 +\ii \lambda_2\,, \quad f= f(z)\,,\label{7_tau}
\end{align}
where $z\equiv x^1+\ii x^2$ is the complex coordinate on the transverse plane 
and $\lambda=C_0+\ii \Exp{-\phi}$ is the axion-dilaton field. 
$f(z)$ and $\lambda(z)$ are holomorphic functions, 
which are determined from the monodromies of 7-branes in $z$-plane. 
Note that the $z$-plane is taken to be a Riemann sphere. 

\begin{figure}[htbp]
\begin{center}
\includegraphics[width=5cm]{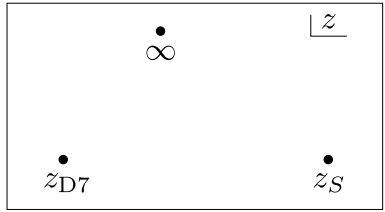}
\caption{The configuration given by Eqs.~\eqref{jtau} and \eqref{simplechoice}. A ${\rm D7}$-brane is at $z=z_{\rm D7}$, an $S$-brane at $z=z_S$, and a 7-brane with the monodromy $T^{-1}\,S$ at $z=\infty$.}
\label{fig:monodromy}
\end{center}
\end{figure}

We choose $f(z)$ and $\lambda(z)$ so that the background includes a ${\rm D7}$-brane, 
and one of the simplest choices is
\begin{align}
 \lambda(z)&= j^{-1}\Bigl(\frac{z_S - z_{\rm D7}}{z-z_{\rm D7}}\Bigr) \,,\label{jtau} \\
 f(z)&= \rho_0^{\frac{1}{3}}\,\eta(\lambda)^2\, (z-z_{\rm D7})^{-\frac{1}{12}}\, 
        (z-z_S)^{-\frac{1}{4}}\,, 
\label{simplechoice}
\end{align}
where $z_{\rm D7}$, $z_S$, and $\rho_0$ are arbitrary complex constants, 
and $\eta$ is the Dedekind eta function,
\begin{align}
 \eta(\lambda)\equiv q^{1/24}\prod_{n=1}^{\infty} (1-q^n)\,, \quad 
 q \equiv \Exp{2\pi \ii \lambda} \,.
\end{align}
$j^{-1}$ is the inverse function of the elliptic modular function $j(\lambda)$ 
which is defined by
\begin{align}
 j(\lambda) \equiv 
 \frac{\bigl(\vartheta_2(\lambda)^8 + \vartheta_3(\lambda)^8 + \vartheta_4(\lambda)^8\bigr)^3}{\eta(\lambda)^{24}}  
\end{align}
with $\vartheta_n(\lambda)$ Jacobi theta functions, and is a one-to-one map from the fundamental region of ${\rm SL}(2,{\mathbb Z})$ 
to the Riemann sphere. 
For simplicity, we take $z_S - z_{\rm D7}$ real in the following. 
The solution given by Eqs.~\eqref{jtau} and \eqref{simplechoice} represents 
a configuration consisting of the following three 7-branes (see Figure \ref{fig:monodromy}):
$$
 \left\{
 \begin{array}{l}
 \text{a ${\rm D7}$-brane, which has the monodromy $T$, at $z=z_{\rm D7}$} \\
 \text{an $S$-brane with the monodromy $S$ at $z=z_S$} \\
 \text{a 7-brane with the monodromy $T^{-1}\,S$ at $z=\infty$}. 
\end{array}
\right.
$$
Here, $T$ and $S$ are the generators of ${\rm SL}(2,{\mathbb Z})$,
\begin{align}
 T= \begin{pmatrix}
 1 & 1\\
 0 & 1
 \end{pmatrix}\,,\qquad 
 S=  \begin{pmatrix}
 0 & -1\\
 1 & 0
\end{pmatrix}\,.
\end{align}
Note that the branes with monodromies $T$, $S$ and $T^{-1}\,S$ correspond to 
the fixed points of the function $j(\lambda)$ under ${\rm SL}(2,\mathbb{Z})$, namely, 
$\lambda=\ii \infty,\ \ii,\ (-1+\ii\sqrt{3})/2$, respectively. 

We can explicitly confirm the presence of these branes 
by taking a limit of approaching each of them and examining the monodromy around it. 
For example, near $z= z_{\rm D7}$, Eq.~\eqref{jtau} becomes\footnote{We used the fact that the relation $j(\lambda)=a z+b$ reduces to $\lambda(z) \sim (\ii/2\pi)\,\log(az+b)$ for large $z$.} 
\begin{align}
 \lambda(z) 
 \overset{\ z\sim z_{\rm D7}}{\sim}
 \frac{\ii}{2\pi}\, \log \Bigl(\frac{z_{S}-z_{\rm D7}}{z - z_{\rm D7}}\Bigr)
 =\frac{\theta}{2\pi} + \frac{\ii}{2\pi}\,\log \Bigl(\frac{r_0}{r}\Bigr)\,,
\label{d7tau}
\end{align}
where we have introduced the polar coordinates,
\begin{align}
 z-z_{\rm D7}\equiv r \Exp{\ii\theta}\,,\qquad 
 z_{S}-z_{\rm D7} \equiv r_0 \in \mathbb{R} \,.\label{polard7}
\end{align}
From Eq.~\eqref{d7tau}, as we go around $z=z_{\rm D7}$, 
i.e.~$\theta \rightarrow \theta+2\pi$, 
we find $\lambda$ shifts as $\lambda\rightarrow \lambda+1$, 
which shows the presence of a ${\rm D7}$-brane at $z=z_{\rm D7}$. 
By using Eq.~\eqref{d7tau}, the factor $\lambda_2\,\lvert f\rvert^2$ 
in Eq.~\eqref{7_metric} near $z=z_{\rm D7}$ is given by 
\begin{align}
 \lambda_2\,\lvert f\rvert^2 
 \overset{\ z\sim z_{\rm D7}}{\sim} \frac{\rho_0^{2/3}}{2\pi r_0^{2/3}}\, \log\Bigl(\frac{r_0}{r}\Bigr)
 =  \frac{1}{2\pi} \, \log\Bigl(\frac{r_0}{r}\Bigr) \,,
\label{tau2f2_d7}
\end{align}
where we have set the arbitrary parameter as $\rho_0 = r_0$. 

Near $z = z_{S}$ (which corresponds to $\lambda =\ii$), 
Eq.~\eqref{jtau} is approximated as
\begin{align}
 j(\lambda)\overset{z\sim z_S}{\sim} 1+ A\,(\lambda-\ii)^2 \label{jtau_s}
\end{align}
with $A$ some real constant. 
By using Eqs.~\eqref{jtau} and \eqref{jtau_s}, we have
\begin{align}
 \lambda \overset{z\sim z_S}{\sim} \ii + \alpha\, (z-z_S)^{1/2} 
 = c\, r^{1/2}\, \cos \Bigl(\frac{\theta}{2} - \delta \Bigr) 
  +\ii \left(1+c\, r^{1/2}\,\sin \Bigl(\frac{\theta}{2} -\delta\Bigr)\right)\,, 
\label{tau_S}
\end{align}
where $\alpha\equiv c \Exp{-\ii\delta}$ is a complex number 
and we have introduced the polar coordinate
\begin{align}
 z-z_S \equiv r\Exp{\ii\theta} \,.\label{polarS}
\end{align}
We can find that as we go around $z=z_{S}$, i.e., $\theta\rightarrow \theta+2\pi$,
$\lambda$ changes as $\lambda \rightarrow -1/\lambda$ (i.e., $S$-duality), 
which indicates the existence of an $S$-brane at $z = z_{S}$. 
The factor $\lambda_2 \,\lvert f\rvert^2$ in Eq.~\eqref{7_metric} 
is approximated near $z=z_{S}$ by
\begin{align}
 \lambda_2 \lvert f\rvert^2\, \overset{z\sim z_S}{\sim} C\, r^{-1/2}\,, 
\label{tau2f2_S}
\end{align}
where $C$ is a constant given by 
$C\equiv r_0^{2/3}\,\bigl\lvert\eta(\ii)^2\,(z_S-z_{\rm D7})^{-\frac{1}{12}}\bigr\rvert^2$.

For $z \sim \infty$, we can similarly check that 
there is a brane which has the monodromy $T^{-1}\,S$. 
The existence of this brane can also be understood as follows. 
A circle which does not enclose any branes has a trivial monodromy, 
but at the same time this circle encloses all the branes existing on the Riemann sphere. 
Therefore, there must be a brane (or a set of branes) which cancels the product 
of monodromies of a ${\rm D7}$- and an $S$-brane, 
which has the monodromy $T$ and $S$ respectively. 

\subsubsection{$5_2^2$- and $T_{34}$-brane solutions and their monodromies}\label{522andI}

In order to obtain $5_2^2$, we take the chain of duality transformations \eqref{chain}
for the finite-energy 7-brane geometry \eqref{7_metric}. 
See appendix \ref{sec:7to522} for the details of the derivation. 
The resulting background which contains a $5^2_2$-brane (in the string frame) is given by
\begin{align}
 \rmd s^2&= \lambda_2\, \lvert f\rvert^2\, \rmd z\, \rmd\bar{z}
           +\frac{\lambda_2}{\lvert\lambda\rvert^2}\, \rmd x_{34}^2 + \rmd x^2_{056789}\,, 
\label{522fromd7-1}\\
 \Exp{2\phi}&= \frac{\lambda_2}{\lvert\lambda\rvert^2}\,,\qquad B_{34} 
             =- \frac{\lambda_1}{\lvert\lambda\rvert^2}\,, \label{522fromd7-2}\\
 \lambda(z) &= j^{-1}\Bigl(\frac{z_S - z_{\rm D7}}{z-z_{\rm D7}}\Bigr) \,, 
\label{522fromd7-3}\\
 f(z)&= r_0^{1/3}\,\eta(\lambda)^2\, (z-z_{\rm D7})^{-\frac{1}{12}}\, (z-z_S)^{-\frac{1}{4}}\,.
\label{522fromd7-4}
\end{align}
Note that  the expressions \eqref{522fromd7-1} and \eqref{522fromd7-2} are valid 
for any choice of $\lambda(z)$ and $f(z)$, 
although we only consider the specific case given in Eqs.~\eqref{522fromd7-3} and \eqref{522fromd7-4}. 
We emphasize that this background is globally defined over the whole $z$-plane. 

Recall that by choosing $\lambda=\lambda_1+\ii\lambda_2$ and $f$ 
as given in Eqs.~\eqref{522fromd7-3} and \eqref{522fromd7-4}, 
we have three 7-branes in the original ${\rm D7}$-brane frame: 
a ${\rm D7}$-brane, an $S$-brane, and a brane with monodromy $T^{-1}\,S$. 
After taking dualities $S\, T_3 \,T_4\, S$, 
these branes are respectively mapped into 
a $5_2^2$ (whose center is at $z=z_{\rm D7}$), a brane which we call a $T_{34}$-brane (at $z=z_S$), 
and a brane (at $z=\infty$) with monodromy $(S\, T_3\, T_4\, S)^{-1}(T^{-1}\,S)(S\, T_3\, T_4\,S)$, 
which is equal to the inverse of the product of the monodromy matrices 
of a $5^2_2$ and a $T_{34}$-brane.    

We can obtain the approximate geometry near the $5_2^2$-brane or the $T_{34}$-brane, 
by rewriting $z$ in terms of $r$ and $\theta$ as in \eqref{polard7} or \eqref{polarS}, 
and by substituting the approximate form of $\lambda$ and $\lambda_2\,\lvert f\rvert^2$ 
near the ${\rm D7}$-brane or near the $S$-brane 
(see Eqs.~\eqref{d7tau}, \eqref{tau2f2_d7}, \eqref{tau_S}, and \eqref{tau2f2_S}) 
into Eqs.~\eqref{522fromd7-1} and \eqref{522fromd7-2}. 
The resulting solution for the near $5_2^2$ geometry 
has the same form as that of Eq.~\eqref{522sol} (which was obtained from the smeared KKM), 
if we identify the parameter $r_0\equiv \lvert z_{S}-z_{\rm D7}\rvert$ 
with the cutoff length $r_c\equiv \mu \Exp{\frac{h}{\gamma}}$. 
The constant $\gamma$, which appears in Eq.~\eqref{522sol}, 
is here given by  $\gamma=1/(2\pi)$. 

Now, we can easily understand the meaning of the cutoff, $r_c$, in Eq.~\eqref{522sol}. 
In the construction from the finite-energy 7-brane solution discussed here, 
the cutoff $r_c$ is given by $r_0$, 
namely the distance of the $5^2_2$ from the neighboring brane 
(which is the $T_{34}$-brane in our case). 
Note that as we have already mentioned, 
we cannot put only a single ${\rm D7}$-brane on $z$-plane 
in order to make the energy density finite. 
Therefore, when we consider a background which contains a $5^2_2$-brane, 
the existence of another neighboring brane is always required 
from the finiteness of  the energy density, 
and the cutoff can always  be interpreted as the distance from the neighboring brane. 

Finally, we examine the monodromies of the $5_2^2$-brane and the $T_{34}$-brane. 
For the background given in Eqs.~\eqref{522fromd7-1} and \eqref{522fromd7-2}, 
the generalized metric of the non-trivially fibered internal $(3,4)$-torus is given by
\begin{align}
 (\mathcal{H}_{AB})
= \frac{1}{\lambda_2} \begin{pmatrix}
 \lvert\lambda\rvert^2\, {\bf 1}& -\lambda_1  \,\boldsymbol{\epsilon} \\
 \lambda_1\,\boldsymbol{\epsilon} & {\bf 1}
\end{pmatrix}\,. \label{gen_met}
\end{align}
The $T$-duality group ${\rm O}(2,2;\mathbb{Z})$ acts on $\mathcal{H}_{AB}$ as
\begin{align}
 \mathcal{H}\rightarrow \Omega^T\, \mathcal{H}\, \Omega\,, \qquad 
 \Omega \in {\rm O}(2,2;\mathbb{Z})\,. \label{o22tf}
\end{align}
Since the $5_2^2$-brane is dual to the ${\rm D7}$-brane in the original frame, 
as we go around the $5_2^2$-brane, 
$\lambda$ in Eq.~\eqref{gen_met} shifts as $\lambda \rightarrow \lambda +1$. 
We can easily find that the change in $\mathcal{H}$ is given 
by the ${\rm O}(2,2;\mathbb{Z})$ transformation \eqref{o22tf} with
\begin{align}
 \Omega(5_2^2)= 
 \begin{pmatrix}
 {\bf 1}& 0 \\
 \boldsymbol{\epsilon} & {\bf 1} 
 \end{pmatrix}\in {\rm O}(2,2;\mathbb{Z}) \,.
\label{522_monodromy} 
\end{align}  
On the other hand, as we go around the $T_{34}$-brane, $\lambda$ in Eq.~\eqref{gen_met} 
changes as $\lambda\rightarrow -1/\lambda$ because 
the $T_{34}$-brane is dual to the $S$-brane in the original frame.  
The associated change in $\mathcal{H}$ is given by 
\begin{align}
 \Omega(T_{34})= 
 \begin{pmatrix}
 0& {\bf 1} \\
 {\bf 1}&0 
 \end{pmatrix}\in {\rm O}(2,2;\mathbb{Z})\,. 
\label{I-brane_monodromy}
\end{align}
Since the the $T_3$- and $T_4$-dualities are given by the matrices
\begin{align}
 T_{3}= \left(
 \begin{smallmatrix}
 0&0&1&0 \\
 0&1&0&0 \\
 1&0&0&0 \\
 0&0&0&1 
 \end{smallmatrix}\right)\,,\qquad
 T_{4}= \left(
 \begin{smallmatrix}
 1&0&0&0 \\
 0&0&0&1 \\
 0&0&1&0 \\
 0&1&0&0 
 \end{smallmatrix}\right)\,, 
\end{align}
we find that the monodromy $\Omega(T_{34})$ is nothing but $T_3 \, T_4$, 
and this is why we call it a $T_{34}$-brane. 

\section{Rotating F-string solutions in non-geometric backgrounds}\label{string_solutions}

In this section, in order to understand 
how objects in string theory behave in a $T$-fold background, 
we explicitly construct classical solutions of a fundamental string rotating 
around a $5_2^2$-brane. 
We also construct the solution of a fundamental string rotating around a $T_{34}$-brane, 
which is given in appendix \ref{sec:T34_string_solution}.

\subsection{Equations of motion}

We start from the Polyakov action in a general background
\begin{align}
 S = -\frac{1}{2}\, \int\rmd^2\sigma\, \bigl[\eta^{\alpha\beta}\, G_{MN}\, \partial_\alpha X^M\, \partial_\beta X^N + \epsilon^{\alpha\beta}\,B_{MN}\,\partial_\alpha X^M\, \partial_\beta X^N\bigr] 
\label{Polyakov_action}
\end{align}
with $\eta^{\alpha\beta} \equiv \begin{scriptsize}\begin{pmatrix} -1&0\cr 0&~1~ \end{pmatrix}\end{scriptsize}$ and $\epsilon^{\alpha\beta}\equiv \begin{scriptsize}\begin{pmatrix} 0 & ~1~\cr -1 &0 \end{pmatrix} \end{scriptsize}$.

As we derived in the previous section, the non-geometric backgrounds considered 
in this paper take the following general form 
(see Eqs.~\eqref{522fromd7-1} and \eqref{522fromd7-2}):
\begin{align}
 &\rmd s^2 = H(r)\,\bigl(\rmd r^2+r^2\,\rmd \theta^2\bigr) 
            + \mathcal{G}(r,\theta) \, \rmd x^2_{34} + \rmd x^2_{056789} \,, \nonumber\\
 & \mathcal{G}(r,\theta)= \frac{\lambda_2}{\lvert\lambda\rvert^2}\,, \qquad 
  B_{34}(r,\theta) = - \frac{\lambda_1}{\lvert\lambda\rvert^2} \,,
\label{general_background}
\end{align}
For the $5^2_2$-brane or the $T_{34}$-brane, 
$\lambda_2\,\lvert f\rvert^2$ depends only on $r$ 
(see Eqs.~\eqref{tau2f2_d7} and \eqref{tau2f2_S}), 
and we thus express it by $H(r)$, i.e., 
$H(r)\equiv\lambda_2\, \lvert f\rvert^2$. 
In this subsection, we only use the fact that  $\lambda$ is a holomorphic function 
in $z$($=\!x^1+\ii x^2$) plane, satisfying the Cauchy-Riemann equations, 
$r\,\partial_r\lambda_1 = \partial_\theta \lambda_2$ and 
$r\,\partial_r\lambda_2 = -\partial_\theta \lambda_1$.
The explicit form of  $\lambda \equiv \lambda_1+\ii \lambda_2$ 
is given in \eqref{d7tau} for near $5_2^2$-brane or in \eqref{tau_S} for near $T_{34}$-brane. 

Now we take the following ansatz
\begin{align}
 X^0(\sigma,\tau)= a\, \tau \,,\quad 
 X^r(\sigma,\tau)=R \,,\quad 
 X^\theta \equiv \Theta(\sigma,\tau)= \omega\,\tau \,,\quad
 X^{i}(\sigma,\tau)=0 \quad (i=5,\dotsc,9)\,, \label{ansatz}
\end{align}
where $a$, $\omega$, and $R$ are some positive constants. 
We leave $X^3(\sigma,\tau)$ and $X^4(\sigma,\tau)$ arbitrary functions to be determined. 
This solution describes an F-string rotating in $z$-plane 
with constant radius $R$ and stretching only in $(3,4)$-torus. 

By defining the complex valued function ${\bf X} \equiv X^3+\ii X^4$, 
the non-zero components of the equations of motion 
and the Virasoro constraints are summarized in the following form 
(see appendix \ref{sec:EOM} for the details):\footnote{Note that $\lvert\dot{\bf X}\rvert^2= (\dot{X}^3)^2+(\dot{X}^4)^2$ and $\lvert{\bf X}'\rvert^2= (X^{3\,\prime})^2+(X^{4\,\prime})^2$. }
\begin{align}
 -\lvert\dot{\bf X}\rvert^2 + \lvert{\bf X}'\rvert^2  
 &= \frac{\omega^2\,\partial_R\bigl(R^2\,H\bigr)}{\lvert\partial_R\lambda\rvert^2}\,
 \bigl[ -2\lambda_1\,\lambda_2\, \partial_R\lambda_1 +(\lambda_1^2-\lambda_2^2)\, \partial_R\lambda_2\bigr]\,,
\label{eq1}\\
 \lvert\dot{\bf X}\rvert^2 + \lvert{\bf X}'\rvert^2
  &= \frac{a^2 - \omega^2\, R^2\,H }{\mathcal{G}} \,,
\label{eq2}\\
 \overline{\dot{\bf X}}\,{\bf X}' 
 &= \ii\, \frac{\omega^2\,\partial_R\bigl(R^2\,H\bigr)}{2\lvert\partial_R\lambda\rvert^2}\,
 \bigl[ 2\lambda_1\,\lambda_2\, \partial_R\lambda_2 +(\lambda_1^2-\lambda_2^2)\, \partial_R\lambda_1 \bigr]\,, 
\label{eq3}\\
 \ddot{\bf X} - {\bf X}'' 
  &= - \frac{\omega}{\mathcal{G}} \, \bigl(\partial_\Theta \mathcal{G} \, \dot{\bf X}
  +\ii\, \partial_\Theta B_{34} \, {\bf X}' \bigr) \,.
\label{eq4}
\end{align}
Here, we defined $\dot{X}^M\equiv \partial_\tau X^M$, $X^{M\,\prime}\equiv \partial_\sigma X^M$ 
and all the quantities are evaluated on the string worldsheet, for example, 
$\partial_\Theta \mathcal{G} \equiv 
\partial_\theta \mathcal{G}(r,\theta)\bigr\rvert_{r=R,\,\theta=\Theta(\tau)}$. 

By plugging Eqs.~\eqref{eq1}--\eqref{eq3} into the following identity,
\begin{align}
 \lvert\overline{\dot{\bf X}}\,{\bf X}'\rvert^2
 =\frac{1}{4} \biggl\{-\bigl( -\lvert\dot{\bf X}\rvert^2 + \lvert{\bf X}'\rvert^2 \bigr)^2
                      +\bigl(\lvert\dot{\bf X}\rvert^2 + \lvert{\bf X}'\rvert^2\bigr)^2\biggr\}\,,
\end{align}
we obtain the relation between the angular velocity $\omega$ and the rotational radius $R$,
\begin{align}
 \Bigl(\frac{\omega^2\,\partial_R\bigl(R^2\,H\bigr)}{2\lvert\partial_R\lambda\rvert^2}\Bigr)^2 
 =  \bigl(a^2- \omega^2\, R^2\,H \bigr)^2.
\end{align}
We must  take the plus or minus sign of the square root of this relation. 
As we will see in the next subsection, which sign we should take 
is determined from the value of $R$, 
namely the distance from the $5_2^2$-brane at the center. 

We rearrange Eqs.~\eqref{eq1}--\eqref{eq3} and summarize the equations to solve:
\begin{align}
 \frac{\omega^2\,\partial_R\bigl(R^2\,H\bigr)}{2\lvert\partial_R\lambda\rvert^2} 
 &=  \pm\bigl(a^2- \omega^2\, R^2\,H \bigr) \,,
\label{plus_minus}\\
 \lvert\dot{\bf X}\rvert^2 &= \frac{a^2- \omega^2\, R^2\,H }{2\lambda_2\,\lvert\partial_R\lambda\rvert} \,\bigl(\lvert\lambda\rvert^2\,\lvert\partial_R\lambda\rvert \mp 2\lambda_1\,\lambda_2\,\partial_R\lambda_1 \pm (\lambda_1^2-\lambda_2^2)\,\partial_R\lambda_2\bigr) \,,
\label{eq1_fin}\\
 {\bf X}' &= \mp\ii\, \frac{2\lambda_1\,\lambda_2\,\partial_R\lambda_2+(\lambda_1^2-\lambda_2^2)\,\partial_R\lambda_1}{\lvert\lambda\rvert^2\,\lvert\partial_R\lambda\rvert \mp2\lambda_1\,\lambda_2\,\partial_R\lambda_1 \pm(\lambda_1^2-\lambda_2^2)\,\partial_R\lambda_2} \, \dot{\bf X}  \,,
\label{eq3_fin}\\
 \ddot{\bf X} - {\bf X}'' 
  &= - \frac{\omega}{\mathcal{G}} \, \bigl(\partial_\Theta \mathcal{G} \, \dot{\bf X}
  +\ii\, \partial_\Theta B_{34} \, {\bf X}' \bigr) \,.
\label{eq4_fin}
\end{align}
Here, Eq.~\eqref{eq1_fin} is obtained from \eqref{eq1} and \eqref{eq2}, 
and Eq.~\eqref{eq3_fin} is obtained by multiplying  \eqref{eq3} with $\dot{\bf X}$ 
and using \eqref{eq1_fin}. 
As we will see below, this overdetermined system of five equations 
for two unknown functions $X^{3,4}(\tau,\sigma)$ indeed has analytical solutions 
for $5^2_2$- and $T_{34}$-branes. 

\subsection{String rotating around $5^2_2$-brane}\label{sec:rotating solution}

In this subsection, we consider the solution of an F-string rotating 
around the $5^2_2$-brane, where the functions $\lambda =\lambda_1+\ii\lambda_2$ 
and $H(r)$ in Eq.~\eqref{general_background} are given by (see Eqs.~\eqref{d7tau} and \eqref{tau2f2_d7})
\begin{align}
 \lambda_1 =\gamma\,\theta \,, \qquad \lambda_2 = H(r) \,, \qquad 
 H(r) = \gamma\,\log(r_c/r)\,.
\label{522_functions}
\end{align}
Depending on the value of $R$ in the ansatz \eqref{ansatz}, 
we have two kinds of solutions as follows.

\paragraph*{\underline{External solution ($\Exp{-\frac{1}{2}}r_c\leq R < r_c$)}\\}

By using the background given in Eq.~\eqref{522_functions}, 
Eqs.~\eqref{eq1_fin}--\eqref{eq4_fin} with upper sign can be written as follows:
\begin{align}
 \lvert\dot{\bf X}\rvert^2 = H\, \bigl(a^2- \omega^2\, R^2\,H \bigr)\,, \quad\ 
 {\bf X}' = \ii\, \kappa\,\tau \, \dot{\bf X} \,,  \quad\ 
 \ddot{\bf X} - {\bf X}'' = \kappa^2\, \tau \, \dot{\bf X} \quad\ 
 \Bigl(\kappa \equiv \frac{\gamma\,\omega}{H} \Bigr)\,, \label{external_eq}
\end{align}
where the value of $\omega$ is determined by using the upper part of Eq.~\eqref{plus_minus} as
\begin{align}
 \omega = \frac{a\,\sqrt{\gamma/2}}{R\,H} \, \sqrt{\frac{H}{\gamma-H}} \,.
\end{align}
Differentiating ${\bf X}' = \ii\, \kappa\,\tau \, \dot{\bf X}$ with respect to $\tau$ or $\sigma$, 
we have 
\begin{align}
 -\kappa^2\, \tau^2\, \ddot{\bf X}- {\bf X}'' = \kappa^2\, \tau\, \dot{\bf X} \,.
\end{align}
By combining this and the last equation in \eqref{external_eq}, we have $\ddot{\bf X} =0$. 
After a short manipulation, 
these equations are solved generally as
\begin{align}
 {\bf X} = {\bf x}_0 + a\sqrt{\frac{H}{2}\frac{\gamma-2H}{\gamma-H}}\, 
                       \Exp{\ii\, \kappa\, (\sigma-\sigma_0)} \, \tau  \,, 
\label{external_solution}
\end{align}
where ${\bf x}_0$ is a constant complex number. 
Generally, this solution describes an open string since the end points of the string 
do not coincide in the spacetime.\footnote{Although we consider the general case here, the closed string is still possible if we adjust the free parameter $a$ to make $\kappa$ even number such that the solution becomes that of a closed string. }
From the physical conditions 
$\lvert\dot{\bf X}\rvert^2 =H(a^2- \omega^2\, R^2\,H ) 
= (a^2\, H/2)\,(\gamma-2H)/(\gamma-H) \geq 0$ and $H>0$, 
we must have $0<H\leq \gamma/2$, that is, the radius $R$ should be within the region; 
\begin{align}
 \Exp{-\frac{1}{2}} r_c \leq R < r_c.
\end{align}
Therefore, this solution cannot exist for the region too close to the $5_2^2$.

\begin{figure}[h]
\begin{center}
\includegraphics[width=14cm]{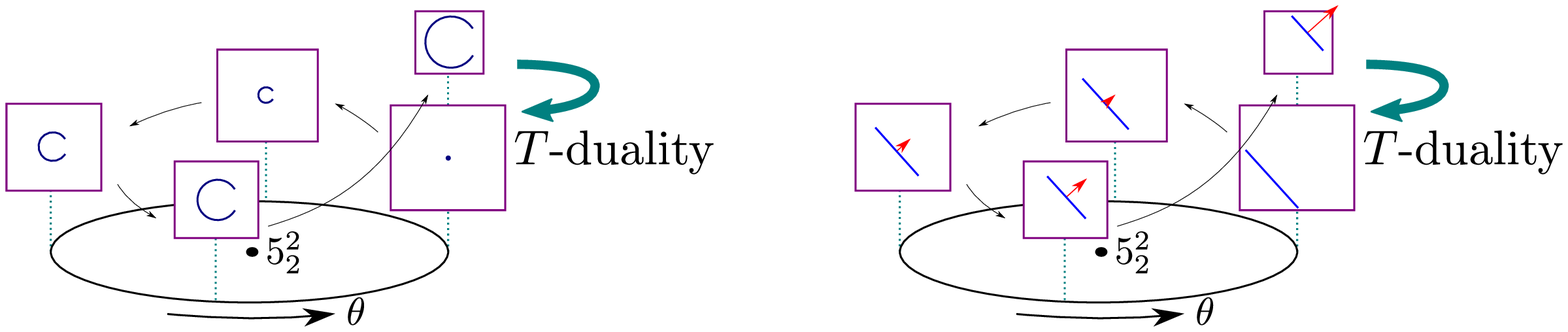}\\
(a) \hskip 7cm (b) ~~~~~~~~~~~~~
\\
\caption{F-string solutions rotating around a $5^2_2$-brane within (a) $\Exp{-\frac{1}{2}}r_c\leq R < r_c$ and (b) $0<R\leq \Exp{-\frac{1}{2}}r_c$. Squares represent internal tori.}
\label{fig:F-string solution}
\end{center}
\end{figure}

The behavior of this solution is shown in Figure \ref{fig:F-string solution} (a). 
The F-string of circular shape expands linearly in $\tau$ 
and does not come back to its original shape when it ends the round trip. 
The dynamics is essentially characterized by 
an ${\rm O}(2,2)$-vector 
\begin{align}
 \bigl(Z^A(\tau,\sigma)\bigr) \equiv {\footnotesize\begin{pmatrix} P_3(\tau,\sigma)\cr 
 P_4(\tau,\sigma) \cr X^{3\,\prime}(\tau,\sigma) \cr X^{4\,\prime}(\tau,\sigma) \end{pmatrix}} \,.
\label{charge density vector}
\end{align}
Here, $P_a$ $(a=3,4)$ are the canonical momenta on the worldsheet, defined by
\begin{align}
 P_a(\tau,\sigma) \equiv G_{aM}\, \dot{X}^M - B_{aM} \, X^{M\,\prime}
 =G_{ab}\, \dot{X}^b - B_{ab} \, X^{b\,\prime}\,.
\label{canonical momenta}
\end{align} 
We call $Z^A(\tau,\sigma)$ the \emph{charge density vector}, 
since its integral over $\sigma$, ${\bf Z}^A(\tau)\equiv \int_0^\pi \rmd \sigma Z^A(\tau,\sigma)$, 
gives the momentum and winding charges for closed string. 
For the solution \eqref{external_solution}, 
the explicit forms of them become as follows (we have set $\sigma_0=0$ for simplicity): 
\begin{align}
 \bigl(Z^A(\tau,\sigma)\bigr)\!=\!{\footnotesize\begin{pmatrix}
 \frac{a}{H}\,\sqrt{\frac{H}{2}\frac{\gamma-2H}{\gamma-H}}\,\cos\kappa \sigma \\
 \frac{a}{H}\,\sqrt{\frac{H}{2}\frac{\gamma-2H}{\gamma-H}}\,\sin\kappa \sigma \\
 -a\sqrt{\frac{H}{2}\frac{\gamma-2H}{\gamma-H}}\,\kappa\,\tau\,\sin\kappa \sigma \\
 a\sqrt{\frac{H}{2}\frac{\gamma-2H}{\gamma-H}}\,\kappa\,\tau\,\cos\kappa \sigma 
\end{pmatrix}}\,,\quad 
 \bigl({\bf Z}^A(\tau)\bigr)\!=\!{\footnotesize\begin{pmatrix}
 \frac{a}{\kappa\,H}\,\sqrt{\frac{H}{2}\frac{\gamma-2H}{\gamma-H}}\, \sin\kappa\pi \\
  -\frac{a}{\kappa\,H}\,\sqrt{\frac{H}{2}\frac{\gamma-2H}{\gamma-H}}\,[\cos\kappa\pi -1] \\
 a\sqrt{\frac{H}{2}\frac{\gamma-2H}{\gamma-H}}\,\tau\, [\cos\kappa \pi -1] \\
 a\sqrt{\frac{H}{2}\frac{\gamma-2H}{\gamma-H}}\,\tau\, \sin\kappa \pi 
\end{pmatrix}}\,.
\end{align}
Thus, the charge density vector $Z^A$ does not return to the initial value 
when the string rotates around the $5_2^2$. 
Note that if $\kappa$ is an even number (i.e., for the solution of a closed string) 
the charge vector, ${\bf Z}^A$, is always zero for this solution. 

We find that this aperiodicity in $Z^A$ is given by the monodromy of $5_2^2$. 
The solution Eq.~\eqref{external_solution} satisfies
\begin{align}
 X^{3\,\prime}=-\gamma\Theta\, P_4\,, \qquad
 X^{4\,\prime}=\gamma\Theta\, P_3\,,
\end{align}
which can be written as 
\begin{align}
 Z^A(\tau,\sigma) = \Omega_{\theta=\omega \tau}^{-1} Z^A(\tau=0,\sigma)\quad\ \text{with} \quad\ 
 \Omega_{\theta=\omega \tau}^{-1}= {\footnotesize
 \begin{pmatrix}
 {\bf 1} & 0 \\
 -\gamma\omega\tau\,\boldsymbol{\epsilon} & {\bf 1}
 \end{pmatrix}} \,.
\label{ZA}
\end{align}
Thus  $Z^A$'s at the same spacetime point are related by 
$\Omega(5^2_2)\in {\rm O}(2,2;{\mathbb Z})$, given by Eq.~\eqref{522_monodromy}. 
As we will see in the next section,  the behavior of $Z^A$ is ``periodic'' 
in a generalized sense, and it is, in fact, ``constant'' along the time evolution. 

The evolution of the charge density vector in \eqref{ZA} 
can also be understood from the worldsheet viewpoint as follows. 
The worldsheet Hamiltonian is given by
\begin{align}
 H= \int \rmd\sigma\, Z^A\,  \mathcal{H}_{AB}\,  Z^B + \text {the other directions}\,, 
\label{ham}
\end{align}
where the first term represents the contribution from $X^3$ and $X^4$. 
If the contribution from the other directions does not include $\tau$, 
which is the case under the ansatz \eqref{ansatz}, 
the first term must be conserved independently. 
Since $\mathcal{H}_{AB}$ varies as in Eq.~\eqref{H_OHO} with $\theta=\omega \tau$, 
the time-dependence of $Z^A$, given in Eq.~\eqref{ZA}, 
is consistent with the fact that the Hamiltonian is conserved.

We comment on the boundary condition of the open string, Eq.~\eqref{external_solution}. 
We can easily check that it satisfies Neumann boundary conditions,
\begin{align}
 \left[G_{MN}\,X^{M\,\prime}-B_{MN}\,\dot{X}^N\right]_{\sigma=0,\,\pi}=0\,,
\label{Neumann}
\end{align}
for $M=3,4,\theta$. 
Thus, in order that this solution makes sense in the perturbative string theory, 
we should fill the whole space spanned by $x^3$, $x^4$ and $\theta$ with some D-branes.  

\paragraph*{\underline{Internal solution ($0 < R \leq \Exp{-\frac{1}{2}} r_c$)}\\}

If we choose the lower sign in Eqs.~\eqref{eq1_fin}--\eqref{eq4_fin}, 
we obtain 
\begin{align}
 \lvert{\bf X}'\rvert^2 = H\, \bigl(a^2- \omega^2\, R^2\,H\bigr)\,, \qquad  
 \dot{\bf X} = \ii\, \kappa\,\tau \, {\bf X}' \,,\qquad 
 \ddot{\bf X} - {\bf X}''  = \ii\, \kappa \, {\bf X}' \,. \label{internal_eq}
\end{align}
Here, the value of $\omega$ is given by Eq.~\eqref{plus_minus} as
\begin{align}
 \omega = (a/H) \, \sqrt{-H'/2R} = (a/HR) \, \sqrt{\gamma/2}  \,.
\end{align}
By using $\dot{\bf X} = \ii\, \kappa\,\tau \, {\bf X}'$, 
we can show that the last equation in \eqref{internal_eq} is equivalent to ${\bf X}'' =0$. 
These can be solved generally as
\begin{align}
 {\bf X} = {\bf x}_0 + \alpha\, \kappa\, \tau^2 - 2 \ii\, \alpha\, \sigma  
 \quad (\alpha\in \mathbb{C})  \label{internal_solution} 
\end{align}
with $\lvert \alpha \rvert^2=(a^2/8)\,(2H-\gamma)$.  
From the physical conditions 
$\lvert{\bf X}'\rvert^2=H(a^2- \omega^2\, R^2\,H) = a^2\, (H-\gamma/2) \geq 0$ and $H>0$, 
we must have $\gamma/2\leq H$, that is,  
the radius $R$ should be within the region; 
\begin{align}
0 < R \leq \Exp{-\frac{1}{2}} r_c. 
\end{align}
Therefore, this solution  exists only for the region close to the $5_2^2$.

The behavior of this solution is shown in Figure \ref{fig:F-string solution} (b). 
The F-string of straight shape moves with constant acceleration in the direction 
perpendicular to itself in internal tori, as it goes around the $5^2_2$-brane. 
Let us see the ${\rm O}(2,2)$-structure of this solution. 
This time, by using the solution Eq.~\eqref{internal_solution} or by the second equation 
in Eqs.~\eqref{internal_eq}, the ${\rm O}(2,2)$-vector is shown to take the following form,
\begin{align}
 \bigl(Z^A(\tau,\sigma)\bigr) =
 {\footnotesize\begin{pmatrix} P_3(\tau,\sigma)\cr 
 P_4(\tau,\sigma) \cr X^{3\,\prime}(\tau,\sigma) \cr X^{4\,\prime}(\tau,\sigma) \end{pmatrix}}
 = {\footnotesize\begin{pmatrix}
 0 \\ 0 \\ 2\,{\rm Im}\,\alpha \\ -2\,{\rm Re}\,\alpha 
\end{pmatrix}}\,.
\end{align}
Therefore, the relation \eqref{ZA} trivially holds. 

\begin{figure}[htbp]
\begin{center}
\includegraphics[width=12cm]{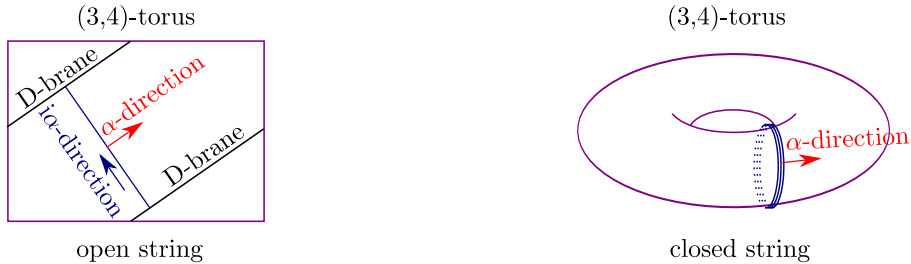}
\caption{The pictorial views of the internal solutions for two kinds of boundary conditions.}
\label{fig:internal}
\end{center}
\end{figure}

We comment on the boundary condition of this internal solution. 
The F-string satisfies Neumann boundary conditions along $\theta$-direction and $\alpha$-direction 
(see the left side of Figure \ref{fig:internal}),
and Dirichlet boundary condition along $\ii \alpha$-direction. 
Thus, in order that this solution makes sense, we should put two parallel D-branes 
extending along $\theta$- and $\alpha$-directions such that they are separate 
in $\ii \alpha$-direction with distance equal to the F-string length.
Alternatively, if the parameter $\alpha$ is the integral multiple of $R_4$ or $\ii R_3$, 
this solution describes a closed string running in the direction of $\alpha$ 
with an integral winding number (see the right side of Figure \ref{fig:internal}). 

\subsection{Evolution of charge density vector and possible string solutions}

We found that the external and the internal solutions satisfy Eq.~\eqref{ZA}, 
which characterizes the time evolution of a string in terms of the monodromy of $5_2^2$. 
In this subsection, we ask about what is the general solution satisfying Eq.~\eqref{ZA}. 
We emphasize that we here are interested only in to what extent the information 
of the monodromy can constrain the rotating string dynamics 
without imposing the other conditions. 
Thus, the obtained solutions are in general off-shell. 

By differentiating Eq.~\eqref{ZA} with respect to $\tau$, 
we have the following equations:
\begin{align}
 {\footnotesize\begin{pmatrix} \dot{P}_3(\tau,\sigma)\cr \dot{P}_4(\tau,\sigma) \cr \gamma \,\omega\, P_4(\tau,\sigma) + \dot{X}^{3\,\prime}(\tau,\sigma) \cr -\gamma\,\omega\, P_3(\tau,\sigma) + \dot{X}^{4\,\prime}(\tau,\sigma) \end{pmatrix}} =0\,.
\label{Lie_transport_eqs}
\end{align}
The last two equations in Eq.~\eqref{Lie_transport_eqs} can be easily solved as 
\begin{align}
 X^3&=-\gamma \,\omega\, \tau\int^\sigma \rmd \sigma'\,P_4(\sigma') +f_3(\tau) +g_3(\sigma) \,,\nonumber\\
 X^4&= \gamma \,\omega\,\tau\int^\sigma \rmd \sigma'\,P_3(\sigma') +f_4(\tau) +g_4(\sigma)\label{x34}
\end{align}
with $f_3(\tau),\ f_4(\tau),\ g_3(\sigma)$ and $g_4(\sigma)$ arbitrary functions. 
Then, the canonical momenta are given by, from Eq.~\eqref{canonical momenta},
\begin{align}
 P_3(\sigma)&= \frac{1}{H}\,\left[ -\gamma\,\omega \int^\sigma\rmd\sigma' \,P_4(\sigma') +f_3'(\tau) + \kappa\,\tau \, g_4'(\sigma) \right] \,,\nonumber\\
 P_4(\sigma)&= \frac{1}{H}\,\left[  \gamma\,\omega \int^\sigma\rmd\sigma' \,P_3(\sigma') +f_4'(\tau) - \kappa\,\tau\, g_3'(\sigma) \right] \,. \label{can mom}
\end{align}
By using $\dot{P}_3(\tau,\sigma)=0= \dot{P}_4(\tau,\sigma)$, 
we can determine the arbitrary functions as follows:
\begin{align}
 f_3 &=c_1\,\kappa\,\tau^2 + c_3\,\gamma\,\omega\,\tau + c_5\,,\quad 
 g_3 =2c_2\,\sigma+c_7 \,,\nonumber\\
 f_4 &=c_2\,\kappa\,\tau^2 - c_4\,\gamma\,\omega\,\tau + c_6 
 \,,\quad g_4 =-2c_1\,\sigma+c_8 \,.
\end{align}
Then, \eqref{can mom} becomes
\begin{align}
 P_3(\sigma) = -\kappa\, \left[ \int^\sigma\rmd\sigma' \,P_4(\sigma') - c_3 \right] \,,\qquad
 P_4(\sigma) = \kappa \,\left[ \int^\sigma\rmd\sigma' \,P_3(\sigma') - c_4 \right] \,,
\end{align}
which can be solved as
\begin{align}
 \int^\sigma\rmd\sigma' \,P_3(\sigma') -c_4 = A\,\sin\kappa(\sigma-\sigma_0) \,,\qquad
 \int^\sigma\rmd\sigma' \,P_4(\sigma') -c_3 = -A\,\cos\kappa(\sigma-\sigma_0) \label{P integral}
\end{align}
with $A$ an arbitrary real number. 
Finally, by substituting  \eqref{P integral} into \eqref{x34},  we have the general solution
\begin{align}
 X^3(\tau,\sigma)&= x^3 + A\,\gamma\,\omega\,\tau\,\cos\kappa(\sigma-\sigma_0) + c_1\,\kappa\,\tau^2 + 2c_2\,\sigma \quad (x^3\equiv  c_5 + c_7) \,,\nonumber \\ 
 X^4(\tau,\sigma)&= x^4 + A\,\gamma\,\omega\,\tau\,\sin\kappa(\sigma-\sigma_0) 
  +c_2\,\kappa\,\tau^2 -2c_1\,\sigma  \quad (x^4\equiv c_6 +c_8)\,. \label{both sol}
\end{align}
Note that both of the external and internal solutions satisfy \eqref{ZA} 
and they actually take the above form. 
Thus, surprisingly enough, the dynamics of the rotating string is determined 
almost completely as given in \eqref{both sol} only from \eqref{ZA}. 
In the next section, we will describe our rotating string solutions in a more natural language. 

\section{The double field theory and the generalized isometry}\label{sec:DFT}

We have constructed the solutions of an F-string rotating around a $5^2_2$-brane 
by solving the equations of motion. 
However, since the background breaks axisymmetry 
only by the action of ${\rm O}(2,2;\mathbb{R})$-transformation (see Eq.~\eqref{H_OHO}), 
we can expect that the geometry has an axial ``isometry'' in a  generalized sense. 
In this section, we show that the $5^2_2$-background indeed has the axial ``isometry'' 
if we describe the background as a doubled geometry. 

In this section, in order to explain what the doubled geometry is,  
we first review the double field theory 
\cite{Hull:2009mi,Hull:2009zb,Hohm:2010jy,Hohm:2010pp,Hohm:2011dv,Hull:2006va}. 
The double field theory has a gauge symmetry 
which is generated by the \emph{generalized Lie derivative}. 
The generalized Lie derivative is a natural extension of the standard Lie derivative, 
and we show that the $5^2_2$-background has an axial isometry 
associated with the generalized Lie derivative. 
In addition, we perform a generalized coordinate transformation 
into a system where the generalized axisymmetry is manifest, 
and show that the charge density vector is constant in $\tau$ in this frame, 
$\partial_\tau Z^I=0$.

\subsection{The double field theory}

Here we briefly review the double field theory. 
In order to consider the $T$-duality, we study the low energy effective theory 
of string theory compactified on a $d$-torus, 
and decompose the local coordinates in the total 10-dimensional space $x^M$ 
($M =0,1,\dotsc,9$) as $(x^M)=(x^\mu,\,x^a)$. 
Here $x^\mu$ are the coordinates in the non-compact spacetime 
($\mu =0,1,\dotsc,9-d$) 
and $x^a$ are the coordinates on the $d$-torus ($a =1,2,\dotsc,d$). 
The low energy effective theory for the NS-NS sector of string theory 
is given by the action
\begin{align}
 S_{\rm eff} 
 = \int \rmd^{10}x\sqrt{-G}\,\Exp{-2\phi}\Bigl(R_G+4\partial_M \phi\, \partial^M\phi 
                               -\frac{1}{12}\,H_{LMN}\,H^{LMN}\Bigr) \,,
\label{NS-NS_action}
\end{align}
where $R_G$ is the Ricci scalar associated with the metric $G_{MN}$, 
and $H_{MNL}$ is the component of the 3-form flux $H=\rmd B^{(2)}$.
While it has symmetries under diffeomorphism $x^M\rightarrow x^M +\delta x^M$ 
and the gauge transformation $B^{(2)}(x)\rightarrow B^{(2)}(x)+\rmd \Lambda (x)$ 
(with $\Lambda(x)$ an arbitrary 1-form), 
the $T$-duality symmetry (which is a non-trivial symmetry in string theory) is not manifest. 

The double field theory is a theory which manifests the $T$-duality invariance 
of the low energy effective theory by \emph{doubling} the coordinates. 
In the double field theory, in addition to the coordinates $x^a$ on the $d$-torus, 
which are associated with momentum excitations $p_a=-\ii\partial_a$, 
we introduce new (periodic) coordinates, $\tilde{x}_a$, 
which are associated with winding excitations 
$w^a=-\ii\tilde{\partial}^a\equiv -\ii\partial/\partial \tilde{x}_a$, 
and deal with these coordinates $(\tilde{x}_a,\,x^a)$ on an equal footing. 
Then the $T$-duality symmetry, i.e., ${\rm O}(d,d;\mathbb{Z})$ symmetry 
can be geometrically understood. 
In particular, as a more radical approach, 
we introduce the doubled coordinates even for non-compact directions \cite{Hohm:2011dv} 
since the double field theory for the NS-NS sector has full ${\rm O}(10,10)$ symmetry 
even when there are no compactification directions. 
We denote the doubled coordinates collectively as 
$y^I=(\tilde{x}_M,\ x^M)$ ($I=\tilde{0},\tilde{1}\dotsc,\tilde{9},0,1,\dotsc,9$). 
The corresponding derivative operators are given by 
$\partial_I =(\tilde{\partial}^M,\, \partial_M)$. 
Here, note that all the fields $G_{MN},\ B_{MN},\ \phi$ are functions of $y^I$. 
We raise and lower the ${\rm O}(10,10)$-indices by using the metric 
\begin{align}
 (\eta_{IJ}) = 
 {\footnotesize\begin{pmatrix}
 0&{\bf 1} \cr
 {\bf 1}&0 
 \end{pmatrix}}  \,.
\end{align}

The action of the double field theory is given by
\begin{align}
 S_{\rm DFT} &= \int \rmd^{20}y \Exp{-2d} \mathcal{R} \,,\\
 \mathcal{R} &\equiv 4 \mathcal{H}^{IJ} \, \partial_I \partial_J d 
  - \partial_I \partial_J \mathcal{H}^{IJ} 
  - 4 \mathcal{H}^{IJ} \, \partial_I d \, \partial_J d 
  + 4 \partial_I \mathcal{H}^{IJ} \, \partial_J d \notag \\
 &\qquad
  + \frac{1}{8} \, \mathcal{H}^{IJ} \, \partial_I \mathcal{H}^{KL} \, \partial_J \mathcal{H}_{KL} 
  - \frac{1}{2} \, \mathcal{H}^{IJ} \, \partial_I \mathcal{H}^{KL} \, \partial_K \mathcal{H}_{JL} \,.
\label{DFT_action}
\end{align}
Here, $\mathcal{H}_{IJ}$ is the generalized metric defined by
\begin{align}
 (\mathcal{H}_{IJ})\equiv {\footnotesize\begin{pmatrix}
  G^{-1} & G^{-1}\,B \\
  -B\, G^{-1} & G-B\, G^{-1}\,B 
 \end{pmatrix}} 
\end{align}
with $G=(G_{MN})$ and $B=(B_{MN})$, and $d$ is the dilaton in the double field theory, 
defined by $\Exp{-2d}\equiv \sqrt{-G}\,\Exp{-2\phi}$.
As the ${\rm O}(10,10)$-indices are properly contracted, 
the above action is manifestly ${\rm O}(10,10)$ invariant. 

We can show that this theory has the following gauge symmetry \cite{Hohm:2010pp}
\begin{align}
 \delta_\xi\mathcal{H}_{IJ} &= 
  \xi^K\partial_K\mathcal{H}_{IJ}+ (\partial_I \xi^K-\partial^K \xi_I)\,\mathcal{H}_{KJ} + (\partial_J \xi^K - \partial^K \xi_J)\,\mathcal{H}_{IK}\,, \nonumber \label{gauge tf} \\
 \delta_\xi(\Exp{-2d}) &= \partial_I (\Exp{-2d}\xi^I) \,,
\end{align}
where $(\xi^I)=(\tilde{\xi}_M,\,\xi^M)$ is a parameter of ${\rm O}(10,10)$-vector. 
By defining the generalized Lie derivatives along an ${\rm O}(10,10)$-vector field $\xi^I$ as
\begin{align}
 \widehat{\mathcal{L}}_\xi V_I & \equiv \xi^J \partial_J V_I + (\partial_I \xi^J - \partial^J \xi_I)\, V_J \,, \nonumber\\
 \widehat{\mathcal{L}}_\xi V^I & \equiv \xi^J \partial_J V^I + (\partial^I \xi_J - \partial_J \xi^I)\, V^J \,,
\label{gen_Lie}
\end{align}
the gauge transformations \eqref{gauge tf} can be written as follows \cite{Hohm:2010pp}:\footnote{Note that in Eq.~\eqref{gauge tf} the dilaton transforms as a scalar density under the gauge transformation. The action of the generalized Lie derivative on tensor densities is defined in a similar manner to that of the usual Lie derivative.}
\begin{align}
 \delta_\xi\mathcal{H}_{IJ}=\widehat{\mathcal{L}}_\xi\mathcal{H}_{IJ} \,,\qquad  
 \delta_\xi(\Exp{-2d})=\widehat{\mathcal{L}}_\xi(\Exp{-2d}) \,.
\end{align}
As well as diffeomorphisms are generated by the Lie derivatives, 
the generalized Lie derivatives generate the diffeomorphisms and the gauge transformations 
$B^{(2)}\rightarrow B^{(2)}+\rmd \Lambda$, 
and more general ${\rm O}(10,10)$-transformations.

The level matching condition $L_0-\bar{L}_0 = N-\bar{N} +\partial_M\,\tilde{\partial}^M=0$ 
in string theory is satisfied by imposing the following strong constraints on
 all supergravity fields and gauge 
parameters 
$A(y)$ and $B(y)$:
\begin{align}
 \partial^I\partial_I A(y) = 0 \,,\qquad \partial^I A(y) \,\partial_I B(y) =0 \,.
\end{align}
These constraints imply that we can always gauge fix such that $\tilde{\partial}^M =0$ 
for all supergravity fields and gauge parameters. 
We can show that the action \eqref{DFT_action} reduces to the action \eqref{NS-NS_action} 
in the gauge, and in this sense, the double field theory has (at least on-shell) 
the same degrees of freedom as the supergravity theory. 
Just as the usual definition of a Killing vector, we define a \emph{generalized Killing vector} 
as a vector $\xi$ that satisfies the following equation:
\begin{align}
 \widehat{\mathcal{L}}_\xi \mathcal{H}_{IJ} = 0 \,.
\end{align}

\subsection{Generalized {axisymmetry} in $5^2_2$ and evolution of the charge density vector}

Let us return to the case of the $5^2_2(56789,34)$ (see Eqs.~\eqref{522sol} and \eqref{gen_metric}), 
in which the internal torus has dimension $d=7$. 
To begin with, although the background does not have an isometry along the axial $\theta$-direction 
in the usual sense, we can easily check that the vector
\begin{align}
 \xi_{\rm axi}
 &\equiv \partial_\theta 
                   + \frac{\gamma}{2}\, \bigl(\tilde{x}_3\partial_4 - \tilde{x}_4\partial_3\bigr) \,
\end{align}
is a generalized Killing vector, by using the matrices
\begin{align}
 &(\partial_A\xi_{\rm axi}^B) = 
 {\footnotesize\begin{pmatrix}
 0&0&0&\gamma/2 \cr
 0&0&-\gamma/2&0 \cr
 0&0&0&0 \cr
 0&0&0&0 
 \end{pmatrix}} \,,\qquad 
 (\partial^A \xi_{\rm axi}{}_B) =
 {\footnotesize\begin{pmatrix}
 0&0&0&0\cr
 0&0&0&0 \cr
 0&\gamma/2&0&0 \cr
 -\gamma/2&0&0&0 
 \end{pmatrix}} \,,
\label{ddxi}
\end{align}
where $A, B=\tilde{3},\tilde{4},3,4$ and other components of 
$\partial_I\xi_{\rm axi}^J$ and $\partial^I \xi_{\rm axi}{}_J$ are zero. 
It should be stressed that this generalized isometry 
could not be found without introducing the doubled geometry. 

{In the following}, we will show that 
the charge density vector $Z^I=(P_M, X^{M'})$ becomes constant in the local coordinate system $\widehat x^{I}$, where 
the components of the generalized metric tensor become 
independent of the axial coordinate $\widehat\theta$.

We consider the following local generalized coordinate transformation,\footnote{Note that this generalized coordinate transformation is consistent with the strong constraint in the sense that the condition (2.7) in \cite{Hohm:2012gk} is satisfied.} 
\begin{align}
  \widehat{x}^{3} = x^3 + (\gamma/2)\,\theta \,\tilde{x}_4\,,\quad 
 \widehat x^{4} = x^4 - (\gamma/2)\,\theta \,\tilde{x}_3\,,\quad 
 \widehat x^{I} = x^I \quad (I\neq 3,4)\,.
 \label{coordinate_transformation}
\end{align}
Then, the Killing vector ${{\xi}_{\rm axi}}$ takes the simple form: $\xi_{\rm axi}=\partial_{\widehat\theta}$. 
Here we have used the tensor transformation law under the generalized coordinate transformation 
proposed in \cite{Hohm:2012gk},
\begin{equation}
 \widehat V_I(\widehat x)=\mathcal{F}_I^{\ J}\, V_{J}(x)\,, \label{vec_tf}
\end{equation}
where the matrix $\mathcal{F}_I^{\ J} $ is defined by
\begin{equation}
 \mathcal{F}_I^{\ J} 
 =\frac{1}{2}\left(\frac{\partial x^K}{\partial \widehat x^{I}}\frac{\partial \widehat x_K}{\partial x_J}
       +\frac{\partial \widehat x_I}{\partial x_K}\frac{\partial x^J}{\partial \widehat x^{K}}\right)\,.
\end{equation}
 In the coordinate $\widehat{x}^{ I}$ given in Eq~\eqref{coordinate_transformation}, 
the transformation matrix $\mathcal{F}_I^{\ J}$ is given by
\begin{align}
 \mathcal{F}_I^{\ J}
 =\footnotesize{\begin{pmatrix}
  1 & 0 & 0 & 0 & 0 & 0 \\
  \frac{\gamma\tilde{x}_4}{2} & 1 & 0 & 0 & 0 & \gamma  \theta  \\
  -\frac{\gamma\tilde{x}_3}{2} & 0 & 1 & 0 & -\gamma  \theta  & 0 \\
  0 & 0 & 0 & 1 & -\frac{\gamma\tilde{x}_4}{2} & \frac{\gamma\tilde{x}_3}{2} \\
  0 & 0 & 0 & 0 & 1 & 0 \\
  0 & 0 & 0 & 0 & 0 & 1
\end{pmatrix}
}\,,
\end{align}
and  the generalized metric $\widehat{{\mathcal H}}_{IJ}$ takes the following form:\footnote{To avoid  notational complication,  we still use  $x^I$ for $r,\ \tilde{3},\ \tilde{4}$, rather than $\widehat{x}^I$, to express $\widehat{\mathcal{H}}_{IJ}$ here.  This notational abuse makes  no confusion since $x^I=\widehat x^I$ for these coordinates.  }
\begin{align}
 &(\widehat{{\mathcal H}}_{IJ})
 =\footnotesize{\begin{pmatrix}
 \frac{1}{R^2 H} & \frac{\tilde{x}_4 \gamma }{2 R^2 H} & -\frac{\tilde{x}_3 \gamma}{2 R^2 H} & 0 & 0 & 0 \\
 \frac{\tilde{x}_4 \gamma}{2 R^2 H} & \frac{\tilde{x}_4^2 \gamma^2}{4 R^2 H}+H & -\frac{\tilde{x}_3 \tilde{x}_4 \gamma^2}{4 R^2 H} & 0 & 0 & 0 \\
 -\frac{\tilde{x}_3 \gamma}{2 R^2 H} & -\frac{\tilde{x}_3 \tilde{x}_4 \gamma^2}{4 R^2 H} & \frac{\tilde{x}_3^2 \gamma^2}{4 R^2 H}+H & 0 & 0 & 0 \\
 0 & 0 & 0 & \frac{(\tilde{x}_3^2+\tilde{x}_4^2) \gamma^2}{4 H}+R^2 H & -\frac{\tilde{x}_4 \gamma}{2 H} & \frac{\tilde{x}_3 \gamma}{2 H} \\
 0 & 0 & 0 & -\frac{\tilde{x}_4 \gamma}{2 H} & \frac{1}{H} & 0 \\
 0 & 0 & 0 & \frac{\tilde{x}_3 \gamma}{2 H} & 0 & \frac{1}{H}
\end{pmatrix}}\,.\label{Hhat}
\end{align}
 Here and below, only $\tilde \theta, \tilde x_3, \tilde x_4, \theta, x^3, x^4$-components are shown in that order. 
Thus, in this coordinate system, 
the generalized metric is  independent of $\widehat \theta$, that is, 
$\widehat{\mathcal{L}}_{\xi_{\rm axi}}\widehat{\mathcal{H}}_{IJ}
 =\partial_{\widehat\theta}\widehat{\mathcal{H}}_{IJ}=0$, 
and the monodromy around the $5_2^2$ becomes trivial.

In order to examine how the charge vector $Z^I$ changes under the coordinate transformation 
Eq~\eqref{coordinate_transformation}, 
we first need to uplift the rotating string solutions $X^M(\sigma,\tau)$
to that of the double sigma model $X^I(\sigma,\tau)$ in the original coordinate system $x^I$. 
The function $\tilde{X}_M(\tau,\sigma)$ in the double sigma model 
is determined by the relation between $X^M$ and $\tilde{X}_M$ \cite{Copland:2011wx},\footnote{In Eq.~\eqref{tildeXdep} with $\alpha=\sigma$, there is, in general,  an additional term in the left-hand side, $\frac{1}{2}\int d^2 \sigma\ \epsilon(\sigma -\sigma') [\partial_M \mathcal{H}_{IJ}\partial_\sigma X^I \partial_\sigma X^J](\sigma')$. However, we can omit this term in \eqref{tildeXdep} because it vanishes for our solution. }
\begin{align}
 \epsilon^{\alpha\beta} \partial_\beta\tilde{X}_M 
 = -(\sqrt{-\gamma}\,G_{MN}\gamma^{\alpha\beta}+B_{MN}\epsilon^{\alpha\beta})\,\partial_\beta X^N \,. 
\label{tildeXdep}
\end{align}
By using this, we find that the solutions $X^I(\sigma,\tau)$ for our rotating strings 
take the following forms:
\begin{align}
 \begin{array}{lcl}
 \text{\underline{external solution}} &\qquad& \text{\underline{internal solution}} \cr
 (X^I) = 
 \footnotesize{\begin{pmatrix}
 \tilde{x}_\theta^0 +\sigma  HR^2\omega \\
 \tilde{x}_3^0+\frac{a}{\gamma\omega}\sqrt{\frac{H}{2}\frac{\gamma-2H}{\gamma-H}}\sin\kappa\sigma \\
 \tilde{x}_4^0-\frac{a}{\gamma\omega}\sqrt{\frac{H}{2}\frac{\gamma-2H}{\gamma-H}}\cos\kappa\sigma \\
 \omega \tau  \\
 x_0^3+a\sqrt{\frac{H}{2}\frac{\gamma-2H}{\gamma-H}}\tau \cos\kappa\sigma \\
 x_0^4+a\sqrt{\frac{H}{2}\frac{\gamma-2H}{\gamma-H}}\tau \sin\kappa\sigma
 \end{pmatrix}}\,,
 &&
 (X^I) = 
 \footnotesize{\begin{pmatrix}
 \tilde{x}_\theta^0+\sigma H R^2\omega \\
 \tilde{x}_3^0+\frac{2{\rm Im}\,\alpha}{H}\tau  \\
 \tilde{x}_4^0-\frac{2{\rm Re}\,\alpha}{H}\tau \\
 \omega  \tau  \\
 x_0^3+\kappa \tau^2\, {\rm Re}\,\alpha +2\sigma\, {\rm Im}\,\alpha  \\
 x_0^4+\kappa\tau^2\,{\rm Im}\,\alpha -2\sigma\, {\rm Re}\,\alpha
 \end{pmatrix}}\,
 \end{array}\,,
\end{align}
where $\tilde x^0_i$ and $x^i_0 ~(i=\theta, 3, 4)$ are integral constants.  
From these solutions, we can obtain the ${\rm O}(10,10)$-vectors, 
$Z^I=\partial_\sigma X^I$ and $Y^I=\partial_\tau X^I$. 
Then, by using the transformation law \eqref{vec_tf}, 
we can obtain the following solutions $\widehat{Z}^I$ in the $\widehat{x}^I$ coordinate;
\begin{align}\begin{array}{lcl}
 \text{\underline{external solution}} &\quad& \text{\underline{internal solution}} \cr
 (\widehat{Z}^I) = 
 \footnotesize{\begin{pmatrix}
 H R^2\omega  +\frac{a^2 (\gamma -2H)}{4\omega (\gamma -H)}
 +\frac{a \kappa}{2 \omega} \sqrt{\frac{H}{2}\frac{\gamma-2H}{\gamma-H}} (\tilde{x}^0_{3} \sin\kappa\sigma -\tilde{x}^0_4\cos\kappa\sigma) \\
 \frac{a}{H} \sqrt{\frac{H}{2}\frac{\gamma-2H}{\gamma-H}} \cos\kappa\sigma \\
 \frac{a}{H} \sqrt{\frac{H}{2}\frac{\gamma-2H}{\gamma-H}} \sin\kappa\sigma \\
 0 \\
 0 \\
 0 
 \end{pmatrix}}\,,
 &&
 (\widehat{Z}^I) = 
 \footnotesize{\begin{pmatrix}
 HR^2\omega \\
 0 \\
 0 \\
 0 \\
 2 {\rm Im}\,\alpha \\
 -2 {\rm Re}\,\alpha 
 \end{pmatrix}}\,.
\end{array}
\end{align}
We thus find that the charge density vectors do not depend on the worldsheet time $\tau$, 
namely $\partial_\tau \widehat{Z}^I=0$. 
This result reflects the fact that the monodromy around $5_2^2$ is trivial 
in the axisymmetric coordinate system $\widehat{x}^I$.


Before closing this section, we comment on some problem 
 associated with the embedding function $X^I(\tau,\sigma)$.
By using Eq.~\eqref{vec_tf}, we can obtain the ${\rm O}(10,10)$-vectors 
$\widehat{Z}^I$ and $\widehat{Y}^I$ in $\widehat{x}^I$-coordinate system 
from the vectors $Z^I $ and $Y^I$ 
in the original coordinate system $x^I$. 
For the external solution, in fact, 
we have $\partial_\tau \widehat{Z}^I\neq \partial_\sigma \widehat{Y}^I$, 
which is expected to be equivalent to 
$\partial_\sigma \partial_\tau\widehat{X}^I -\partial_\tau \partial_\sigma\widehat{X}^I\neq 0$. 
If it is the case, we have to conclude that, for the external solution, 
the embedding map $\widehat{X}^I(\sigma,\tau)$ itself cannot be obtained in the $\widehat x^{I}$-coordinates.
On the other hand, for the internal solution, the relation 
$\partial_\tau \widehat{Z}^I =  \partial_\sigma  {\widehat{Y}}^I$ does hold, 
and $\widehat{X}^I$ can be obtained by assuming $\widehat{Z}^I=\partial_\sigma \widehat{X}^I$ and $\widehat{Y}^I=\partial_\tau \widehat{X}^I$. 
However, the resulting configuration $\widehat{X}^I$ is not the same as the one 
that can obtained by Eq.~\eqref{coordinate_transformation} from the original solution $X^I(\tau,\sigma)$. 
We have little understanding of such strange situations, 
and leave further investigation to future work.

\section{Conclusion and discussions}\label{sec:conclusion}

We have constructed a globally defined $T$-fold background  
which contains a $5^2_2$ and other 7-branes. 
Near the center of $5_2^2$, the geometry approaches 
to that constructed in \cite{deBoer:2010ud}, 
which includes a cutoff parameter and is not globally defined. 
We have then shown that the cutoff parameter is {to be} interpreted 
as the distance of the $5^2_2$ from the other neighboring 7-branes. 
The monodromies of these 7-branes in the background were also studied. 

One of the main purposes in this paper is 
to understand how objects in string theory behave in $T$-fold backgrounds. 
In particular, we have constructed the explicit solutions of 
a fundamental string rotating around $5^2_2$, and have found that
the time evolution of the charge density 
is given by $\Omega_{\theta=\omega \tau}$ as in \eqref{ZA}. 
Conversely, the relation \eqref{ZA} by itself essentially determines 
the possible on-shell behavior of a rotating F-string. 

In order to interpret the monodromy action more geometrically, 
we have described the $5_2^2$ as a doubled geometry. 
We have found the generalized Killing vector {$\xi_{\rm axi}$} which generates the axial rotation 
(accompanied by the non-trivial transformation in the $(3,4)$-torus). 
Moreover, we have found that 
the charge density vectors for our string solution 
are invariant under the time evolutions, $\partial_\tau \widehat Z^I = 0$, 
in the coordinate system where the generalized isometry is manifest, 
$\partial_{\widehat\theta}\widehat{\mathcal H}_{IJ}=0$. 
In this paper we have provided such a direct application of the rather formal 
and still growing framework of the double field theory 
to a simple and familiar example as the motion of F-strings. 

As we pointed out in the Introduction, 
the codimension-2 branes with non-trivial monodromies 
examined in this paper can be regarded as a realization of Alice strings 
in the context of string theory. 
In the original theory of Alice string \cite{Schwarz:1982ec}, 
the monodromy group is simply ${\mathbb Z}_2$. 
Nevertheless, the Alice string brings curious notions 
such as non-locally conserved charges 
(called \emph{Cheshire charges} \cite{Alford:1990mk,Alford:1990ur}). 
It is then quite interesting to study what phenomena happen in string theory, 
which has highly non-trivial $U$-duality symmetry 
and contains various branes as charged objects. 
For example, if we consider a ${\rm D1}$-brane rotating around a ${\rm D7}$-brane, 
we expect, because of the ${\rm D7}$-brane monodromy, 
that an F-string charge is induced on the ${\rm D1}$-brane along the way. 
In order to keep charge conservation law, the induced charge 
should be absorbed from the background. 
To examine such a process, we expect that we can apply 
an analysis given in \cite{Gregory:1997te} to our non-geometric case. 
Although we have considered classical solutions of an F-string in this paper, 
the authors in \cite{Gregory:1997te} considered an off-shell trajectory 
of an F-string in a KKM background. 
In this background, 
there is a continuous trajectory where an initially winding string becomes unwound, 
and the background was found to absorb the winding charge. 
It would be interesting to perform a similar analysis for non-geometric background 
such as $5_2^2$ and examine the charge transition between a probe and the background explicitly. 
Finally, we must mention a subtlety of the definition of charges in IIA or IIB supergravity. 
As is discussed in \cite{Marolf:2000cb}, in theories of supergravity with Chern-Simons terms, 
we can define three inequivalent notions of charges. 
Therefore, in the process of a probe going around a brane with non-trivial monodromy, 
it is important to clarify which type of the charges varies.\footnote{We would like to thank Masaki Shigemori for helpful comments on this point.}

\section*{Acknowledgments}

The authors would like to thank Satoshi Iso, Tetsuji Kimura, Muneto Nitta, 
Masaki Shigemori and Shinji Shimasaki for helpful discussions.  
This work was supported by the Japan Society for the Promotion of Science (JSPS) 
(Grant Nos.~21$\cdot$951, 24$\cdot$1681, and 21$\cdot$1105).  
This work was also supported in part by the Grant-in-Aid for the Global COE Program 
``The Next Generation of Physics, Spun from Universality and Emergence'' 
from the Ministry of Education, Culture, Sports, Science and Technology 
(MEXT) of Japan.

\appendix

\section{$5_2^2$ from D7-brane}\label{sec:7to522}

In section \ref{522andI}, we derived the background of $5_2^2(56789,34)$ 
from that of ${\rm D7}(3456789)$-brane by taking the following sequence of dualities, $S\,T_3\,T_4\,S$;
$$
 {\rm D7}(3456789) \overset{S}{\rightarrow}
 {\rm NS7}(3456789) \overset{T_3}{\rightarrow}
 6^1_3 (456789,3) \overset{T_4}{\rightarrow}
 5^2_3(56789,34) \overset{S}{\rightarrow}
 5^2_2(56789,34) \,.
$$
In this appendix, we write down the backgrounds obtained 
in each step of the dualities in terms of the parameters appearing 
in the ${\rm D7}$-brane solution \eqref{7_metric} and \eqref{7_tau}, 
i.e.~$\lambda=\lambda_1+ \ii \lambda_2$ and $f=f(z)$. 
Note that we express the metrics for them in string frame.

By taking $S$-duality for ${\rm D7}(3456789)$-brane, 
we obtain the ${\rm NS7}(3456789)$-brane background;
\begin{align}
 \rmd s^2 ({\rm NS7}) &= \Exp{\frac{\phi}{2}}\, \rmd x_{03456789}^2 
  +\Exp{\frac{\phi}{2}} \lambda_2 \lvert f\rvert^2\, \rmd x_{12}^2 \,, \nonumber\\ 
\Exp{\phi}& = \biggl(\frac{\lambda_2}{\lvert\lambda\rvert^2}\biggr)^{-1}\,,\qquad 
C^{(0)}=  -\frac{\lambda_1}{\lvert\lambda \rvert^2}.
\end{align}

By taking $T_3$, the ${\rm NS7}(3456789)$-brane becomes $6_3^1(456789,3)$;
\begin{align}
\rmd s^2 (6_3^1)&= \Exp{\frac{2}{3}\phi} \rmd x_{0456789}^2 + \Exp{-\frac{2}{3}\phi} (\rmd x^3)^2 +\Exp{\frac{2}{3}\phi} \lambda_2 \lvert f\rvert^2\, \rmd x_{12}^2 \,,\nonumber\\
 \Exp{\phi}&= \biggl(\frac{\lambda_2}{\lvert\lambda\rvert^2}\biggr)^{-3/4}\,,\qquad 
 C^{(1)}= -\frac{\lambda_1}{\lvert\lambda \rvert^2}\,\rmd x^3\,,
\end{align}
where $C^{(1)}$ is RR 1-form.

By further taking $T_4$, the $6_3^1$ becomes $5_3^2(56789,34)$;
\begin{align}
 \rmd s^2(5_3^2) &= \Exp{\phi} \rmd x_{056789}^2 + \Exp{-\phi} \rmd x_{34}^2 +\Exp{\phi} \lambda_2 \lvert f\rvert^2\, \rmd x_{12}^2\,,\nonumber\\
 \Exp{\phi}&= \biggl(\frac{\lambda_2}{\lvert\lambda\rvert^2}\biggr)^{-1/2}\,,\qquad 
 C^{(2)}= -\frac{\lambda_1}{\lvert\lambda\rvert^2}\, \rmd x^3\wedge \rmd x^4\,,
\end{align}
where $C^{(2)}$ is RR 2-form.

Finally, taking $S$-duality for the $5_3^2(56789,34)$, 
we obtain $5_2^2(56789,34)$ background, 
which is given in Eqs.~\eqref{522fromd7-1} and \eqref{522fromd7-2}.

\section{Equations of motion of fundamental string}\label{sec:EOM}

The equation of motion of a fundamental string is given by ${\rm E}_M\equiv -(2/\sqrt{-\eta})\,\delta S/\delta X^M=0$, which can be written down as 
\begin{align}
 {\rm E}_M 
  =&~ 2 G_{MN}\,\bigl(\ddot{X}^N- X^{N\prime\prime}\bigr)
   + 2\Gamma_{MNL}\,\bigl(\dot{X}^N\,\dot{X}^L - X^{N\,\prime}\, X^{L\,\prime}\bigr)
   + 2 H_{MNL} \,\dot{X}^{N}\,X^{L\,\prime}=0 \,,
\end{align}
where $H_{MNL}$ are the components of the 3-form flux $H=\rmd B^{(2)}$.

In the general background \eqref{general_background}, by using the ansatz \eqref{ansatz}, 
the non-zero components of $E_M$ 
are given as follows:
\begin{align}
 {\rm E}_R =&~ \partial_R \mathcal{G}\,\bigl[-(\dot{X}^3)^2-(\dot{X}^4)^2+(X^{3\,\prime})^2+(X^{4\,\prime})^2\bigr]  \nonumber\\
 &~  + 2\partial_R B_{34}\,\bigl(\dot{X}^3\,X^{4\,\prime}-X^{3\,\prime}\,\dot{X}^4 \bigr) - \omega^2\,\partial_R\bigl(R^2\,H\bigr) \,,\nonumber\\
 {\rm E}_{\Theta} =&~ \partial_{\Theta} \mathcal{G}\,\bigl[-(\dot{X}^3)^2-(\dot{X}^4)^2+(X^{3\,\prime})^2+(X^{4\,\prime})^2\bigr] \nonumber\\
 &~ + 2\partial_{\Theta} B_{34}\,\bigl(\dot{X}^{3}\,X^{4\,\prime}-X^{3\,\prime}\,\dot{X}^4 \bigr) \,, \\
 {\rm E}_{3} =&~ 2\bigl[\omega\, \partial_\Theta \mathcal{G} \, \dot{X}^3
                        + \mathcal{G}\, \bigl(\ddot{X}^3 - X^{3\,\prime\prime} \bigr)
                        - \omega\, \partial_\Theta B_{34} \, X^{4\,\prime} \bigr] \,,\nonumber\\
 {\rm E}_{4} =&~ 2\bigl[\omega\,\partial_\Theta \mathcal{G} \, \dot{X}^4
                        + \mathcal{G}\, \bigl(\ddot{X}^4 - X^{4\,\prime\prime} \bigr)
                        + \omega\,\partial_\Theta B_{34} \,X^{3\,\prime} \bigr] \,.\nonumber
\end{align}
The Virasoro constraints are given by
\begin{align}
 0 &= \dot{X}^3\,X^{3\,\prime} + \dot{X}^4\,X^{4\,\prime} \,,\nonumber\\
 0 &= \mathcal{G}\,\bigl[(\dot{X}^3)^2+(\dot{X}^4)^2+(X^{3\,\prime})^2
       +(X^{4\,\prime})^2\bigr] + \omega^2\, R^2\,H  -a^2 \,.
\end{align}
By introducing the complex valued function ${\bf X} \equiv X^3+\ii X^4$, 
the above equations are  summarized in the form given in 
Eqs.~\eqref{eq1}--\eqref{eq4}.

\section{Rotating F-string solution around $T_{34}$-brane}\label{sec:T34_string_solution}

The background near the $T_{34}$-brane is given by the approximation
\begin{align}
 &\lambda_1 = c\,\sqrt{r} \cos (\theta/2 -\delta) \,,\qquad 
  \lambda_2 = 1 +c\,\sqrt{r} \sin (\theta/2 -\delta) \,,\nonumber\\ 
 &\mathcal{G}=\frac{1}{2} \left(1+\frac{1-c^2\,r}{1+c^2\,r +2 c\, \sqrt{r}\, \cos\bigl(\frac{\omega\,\tau}{2}- \delta -\frac{\pi}{2}\bigr)}\right) \,,\\
 &B_{34}= \frac{c\,\sqrt{r}\,\sin\bigl(\frac{\omega\,\tau}{2}- \delta -\frac{\pi}{2}\bigr)}{1+c^2\,r +2 c\, \sqrt{r}\, \cos\bigl(\frac{\omega\,\tau}{2}- \delta -\frac{\pi}{2}\bigr)} \,,\quad  
 H= C\, r^{-1/2}\,. \nonumber
\end{align}

The value of $\omega$ is determined from Eq.~\eqref{plus_minus} as\footnote{We note that  the negative sign in Eq.~\eqref{plus_minus} is not allowed for small $R$, since for the negative sign, $\omega^2$ becomes negative, $\omega^2 = \frac{c\, a^2}{-3\,C\,R+c\,C\, R^{3/2}} \sim -\frac{c\, a^2}{3\,C\,R} \,.$}
\begin{align}
 \omega^2 = \frac{c\, a^2}{3\,C\,R+c\,C\, R^{3/2}} \,.
\end{align}

In this near $T_{34}$-brane background, Eqs.~\eqref{eq1_fin}--\eqref{eq4_fin} take the following form:
\begin{align}
 \lvert\dot{\bf X}\rvert^2 &= \frac{\bigl(1-c\,\sqrt{R}\bigr)^2\, \bigl(a^2-H\,R^2\,\omega^2\bigr)\, \sin^2\bigl(\frac{\omega\,\tau}{4}-\frac{\delta}{2}-\frac{\pi}{4}\bigr)}{1+c\, \sqrt{R}\, \cos\bigl(\frac{\omega\,\tau}{2}- \delta -\frac{\pi}{2}\bigr)} \,, \nonumber\\
 \dot{\bf X} &=  \ii\frac{\bigl(1-c\,\sqrt{R}\bigr)\, \tan\bigl(\frac{\omega\,\tau}{4}-\frac{\delta}{2}-\frac{\pi}{4}\bigr)}{1+c\, \sqrt{R}} \, {\bf X}'  \,, \\
 \ddot{\bf X}-{\bf X}'' &= -\frac{\ii}{2}\, \frac{c\,\omega\, \sqrt{R}}{1+c\, \sqrt{R}\, \cos\bigl(\frac{\omega\,\tau}{2}- \delta -\frac{\pi}{2}\bigr)} \, {\bf X}' \,.\nonumber
\end{align}
Since it is difficult to solve these equations for generic values of $R$, 
we restrict ourselves to the case, $c^2 R\ll 1$. 
Then, we obtain the following approximate equations:
\begin{align}
 &\lvert\dot{\bf X}\rvert^2 = a^2 \, \sin^2\Bigl(\frac{\omega\,\tau}{4}-\frac{\delta}{2}-\frac{\pi}{4}\Bigr) \,, \quad 
 \dot{\bf X} = \ii\,\tan\Bigl(\frac{\omega\,\tau}{4}-\frac{\delta}{2}-\frac{\pi}{4}\Bigr) \, {\bf X}'  \,, \qquad
 \ddot{\bf X}-{\bf X}'' =  0 \,.
\end{align}
Then, the solution is given by
\begin{align}
 {\bf X} ={\bf x}_0 + \frac{4a}{\omega}\,\cos\Bigl(\frac{\omega\,\tau}{4}-\frac{\delta}{2}-\frac{\pi}{4}\Bigr)\,\Exp{\ii\omega\,(\sigma-\sigma_0)/4} \,,
\end{align}
where $\omega = a\,\sqrt{c/3\,C\,R}$.

\end{document}